\documentclass[12pt,letterpaper]{article}

\usepackage{xcolor, colortbl}
\usepackage{amssymb, amsmath, amsthm}
\usepackage{graphicx}
\usepackage{natbib}
\usepackage{float}
\usepackage{setspace}
\usepackage{algorithm}
\usepackage{algorithmic}

\definecolor{lightgray}{gray}{0.9}

\newcommand{\di}{\text{d}}
\newcommand{\bX}{\mathbf{X}}
\newcommand{\bx}{\mathbf{x}}

\newcommand{\by}{\mathbf{y}}
\newcommand{\bZ}{\mathbf{Z}}

\newcommand{\bK}{\mathbf{K}}

\newcommand{\bV}{\mathbf{V}}
\newcommand{\bW}{\mathbf{W}}
\newcommand{\ba}{\mathbf{a}}
\newcommand{\bc}{\mathbf{c}}

\newcommand{\bh}{\mathbf{h}}

\newcommand{\bA}{\mathbf{A}}
\newcommand{\bB}{\mathbf{B}}

\newcommand{\bH}{\mathbf{H}}

\newcommand{\bw}{\mathbf{w}}
\renewcommand{\epsilon}{\varepsilon}
\renewcommand{\hat}{\widehat}
\renewcommand{\tilde}{\widetilde}

\newcommand{\distn}[1]{\mathcal{#1}}

\newcommand{\Em}{\mathbb E}
\newcommand{\Pm}{\mathbb P}

\newcommand{\gvn}{\,|\,}
\newcommand{\e}{\text{e}}

\newcommand{\vect}[1]{\boldsymbol #1}
\newcommand{\vtheta}{\vect{\theta}}

\newcommand{\vbeta}{\vect{\beta}}
\newcommand{\vmu}{\vect{\mu}}

\newcommand{\vepsilon}{\vect{\epsilon}}

\newcommand{\vOmega}{\vect{\Omega}}
\newcommand{\vSigma}{\vect{\Sigma}}

\newcommand{\vpsi}{\vect{\psi}}
\newcommand{\vgamma}{\vect{\gamma}}

\newcommand{\vkappa}{\vect{\kappa}}
\newcommand{\valpha}{\vect{\alpha}}

\DeclareMathOperator{\diag}{diag}
\newcommand{\matlab}{\mathrm{M}\mathrm{{\scriptstyle ATLAB}}}

\bibpunct{(}{)}{;}{a}{,}{,}

\begin{document}

\title{Large Hybrid Time-Varying Parameter VARs}

\author{Joshua C.C. Chan\thanks{This paper has benefited from the constructive comments and suggestions from many people. I particularly thank Mark Bognanni, Todd Clark, Francesco Corsello, Francis Diebold, Luis Uzeda Garcia, Jaeho Kim, Kurt Lunsford, Michael McCracken, Elmar Mertens, Jouchi Nakajima, Davide Pettenuzzo, Frank Schorfheide, Le Wang, Benjamin Wong and Saeed Zaman, as well as conference and seminar participants at the 27th Annual Symposium of the SNDE, Applied Time Series Econometrics Workshop at the Federal Reserve Bank of St. Louis, the Deutsche Bundesbank, the Federal Reserve Bank of Cleveland, the 5th Hitotsubashi Summer Institute, Oklahoma University and University of Pennsylvania. All remaining errors are, of course, my own.} \\
 {\small Purdue University}
}

\date{This version: May 2022\\
First version: September 2019
}

\maketitle

\onehalfspacing


\begin{abstract}

\noindent Time-varying parameter VARs with stochastic volatility are routinely used for structural analysis and forecasting in settings involving a few endogenous variables. Applying these models to high-dimensional datasets has proved to be challenging due to intensive computations and over-parameterization concerns. We develop an efficient Bayesian sparsification method for a class of models we call hybrid TVP-VARs---VARs with time-varying parameters in some equations but constant coefficients in others. Specifically, for each equation, the new method automatically decides whether the VAR coefficients and contemporaneous relations among variables are constant or time-varying. Using US datasets of various dimensions, we find evidence that the parameters in some, but not all, equations are time varying. The large hybrid TVP-VAR also forecasts better than many standard benchmarks.

\bigskip




\end{abstract}

\thispagestyle{empty}

\newpage

\section{Introduction}

Time-varying parameter vector autoregressions (TVP-VARs) developed by \citet{CS01, CS05} and \citet{Primiceri05} have become the workhorse models in empirical macroeconomics. These models are flexible and can capture many different forms of structural instabilities and the evolving nonlinear relationships between the dependent variables. Moreover, they often forecast substantially better than their homoskedastic or constant-coefficient counterparts, as shown in papers such as \citet{clark11}, \citet*{DGG13}, \citet{KK13}, \citet{CR15} and \citet{CP16}. In empirical work, however, their applications are mostly limited to modeling small systems involving only a few variables because of the computational burden and over-parameterization concerns.

On the other hand, large VARs that use richer information have become increasingly popular due to their better forecast performance and more sensible impulse-response analysis, as demonstrated in the influential paper by \citet*{BGR10}. There is now a rapidly expanding literature that uses large VARs for forecasting and structural analysis. Prominent examples include \citet*{CKM09}, \citet{koop13}, \citet{BGMR13}, \citet*{CCM15}, \citet{ER17} and \citet{MW19}. Since there is a large body of empirical evidence that demonstrates the importance of accommodating time-varying structures in small systems, there has been much interest in recent years to build TVP-VARs for large datasets. While there are a few proposals to build large constant-coefficient VARs with stochastic volatility \citep[see, e.g.,][]{CCM16, CCM19, KH18, chan20, chan21}, the literature on large VARs with time-varying coefficients remains relatively scarce. 

We propose a class of models we call hybrid TVP-VARs---VARs in which some equations have time-varying coefficients, whereas the coefficients are constant in others. More precisely, we develop an efficient Bayesian shrinkage and sparsification method that automatically decides, for each equation, (i) whether the VAR coefficients are constant or time-varying, and (ii) whether the parameters of the contemporaneous relations among variables are constant or time-varying. Given the importance of time-varying volatility, all equations feature stochastic volatility. Our framework nests many popular VARs as special cases, ranging from a constant-coefficient VAR with stochastic volatility on one end of the spectrum to the flexible but highly parameterized TVP-VARs of \citet{CS05} and \citet{Primiceri05} on the other end. More importantly, our framework also includes many hybrid TVP-VARs in between the extremes, allowing for a more nuanced modeling approach of the time-varying structures. 

To formulate these large hybrid TVP-VARs, we use a reparameterization of the standard TVP-VAR in \citet{Primiceri05}. Specifically, we rewrite the TVP-VAR in the structural form in which the time-varying error covariance matrices are diagonal. Hence, we can treat the structural TVP-VAR as a system of $n$ unrelated TVP regressions and estimate them one by one. This reduces the dimension of the problem and can substantially speed up computations. This approach is similar to the equation-by-equation estimation approach in \citet*{CCM19} that is designed for the reduced-form parameterization. But since under our parameterization there is no need to obtain the `orthogonalized' shocks at each iteration as in \citet*{CCM19}, the proposed approach is substantially faster. Moreover, under our parameterization the estimation can be parallelized to further speed up computations. This structural-form parameterization, however, raises the issue of variable ordering, that is, the assumed order of the variables might affect the model estimates compared to a standard reduced-form TVP-VAR. We investigate this issue empirically and find that the variability of the estimates from this structural-form parameterization is comparable to that of the TVP-VAR of \citet{Primiceri05}. 

Next, we adapt the non-centered parameterization of the state space model in \citet{FSW10} to our structural TVP-VAR representation. Further, for each equation we introduce two indicator variables, one determines whether the VAR coefficients are time-varying or constant, while the other controls whether the elements of the impact matrix are time-varying or not. Hence, each vector $\vgamma\in\{0,1\}^{2n}$, where $n$ is the number of endogenous variables, characterizes a hybrid TVP-VAR with a particular form of time variation. By treating these indicators as parameters to be estimated, we allow the data to determine the appropriate time-varying structures, in contrast to typical setups where time variation is assumed. The proposed approach therefore is not only flexible---it includes many state-of-the-art models routinely used in applied work as special cases---it also induces parsimony to ameliorate over-parameterization concerns. This data-driven hybrid TVP-VAR can also be interpreted as a Bayesian model average of $2^{2n}$ hybrid TVP-VARs with different forms of time variation, where the weights are determined by the posterior model probabilities $p(\vgamma\gvn\by)$. It follows that forecasts from such a model can be viewed as a forecast combination of a wide variety of hybrid TVP-VARs. 

The estimation is done using Markov chain Monte Carlo (MCMC) methods. Hence, in contrast to earlier attempts to build large TVP-VARs, our approach is fully Bayesian and is exact---it simulates from the exact posterior distribution. There are, however, a few challenges in the estimation. First, the dimension of the model is large and there are thousands of latent state processes---time-varying coefficients and stochastic volatilities---to simulate. To overcome this challenge, in addition to using the equation-by-equation estimation approach described earlier, we also adopt the precision sampler of \citet{CJ09} to draw both the time-invariant and time-varying VAR coefficients, as well as the stochastic volatilities. In our high-dimensional setting the precision sampler substantially reduces the computational cost compared to conventional Kalman filter based smoothers. A second challenge in the estimation is that the indicators and the latent states enter the likelihood multiplicatively. Consequently, it is vital to sample them jointly; otherwise the Markov chain is likely to get stuck. We therefore develop algorithms to sample the indicators and the latent states jointly. 

Using US datasets of different dimensions, we find evidence that the VAR coefficients and elements of the impact matrix in some, but not all, equations are time varying. In particular, in a formal Bayesian model comparison exercise, we show that there is overwhelming support for the (data-driven) hybrid TVP-VAR relative to a few standard benchmarks, including a constant-coefficient VAR with stochastic volatility and a full-fledged TVP-VAR in which all the VAR coefficients and error covariances are time varying. We further illustrate the usefulness of the hybrid TVP-VAR with a forecasting exercise that involves 20 US quarterly macroeconomic and financial variables. We show that the proposed model forecasts better than many benchmarks. These results suggest that using a data-driven approach to discover the time-varying structures---rather than imposing either constant coefficients or time-varying parameters---is empirically beneficial.

This paper contributes to the budding literature on developing large TVP-VARs. Earlier papers include \citet{KK13, KK18}, who propose fast methods to approximate the posterior distributions of large TVP-VARs. \citet{BV18} and \citet{GH18} consider large VARs with only time-varying intercepts. \citet*{CES20} model the time-varying coefficients using a factor-like reduced-rank structure, whereas \citet{HKO19} develop a method that first shrinks the time-varying coefficients, followed by setting the small values to zero. As mentioned above, our estimation approach is exact and fully Bayesian, and the modeling framework is more flexible than many of those in earlier papers. There is also a growing literature on alternative, non-likelihood based approaches. Examples include \citet{GKP13} and \citet{Petrova19} that allow for the estimation of large TVP-VARs without imposing the Cholesky-type stochastic volatility, and hence they avoid the ordering issue. Nevertheless, one main advantage of the likelihood-based approach taken in this paper is that it is flexible and modular. In particular, it is straightforward to incorporate additional useful features into the proposed hybrid model, such as more sophisticated static and dynamic shrinkage priors for VARs \citep{pruser21,chan21} or more flexible error distributions to deal with outliers \citep{CCMM21,BH21}.


The rest of the paper is organized as follows. We first introduce the proposed modeling framework in Section~\ref{s:models}. In particular, we discuss how we combine a reparameterization of the reduced-form TVP-VAR and the non-centered parameterization of the state space model to develop the hybrid TVP-VARs. We then describe the shrinkage priors and the posterior sampler in Section~\ref{s:estimation}. 
It is followed by a Monte Carlo study in Section~\ref{s:MC} that demonstrates that the proposed methodology works well and can select the correct time-varying or time-invariant structure. The empirical application is discussed in detail in Section~\ref{s:application}. Lastly, Section~\ref{s:conclusions} concludes and briefly discusses some future research directions.

\section{Hybrid TVP-VARs} \label{s:models}

We first introduce a class of models we call hybrid time-varying parameter VARs: VARs in which some equations have time-varying coefficients, whereas coefficients in other equations remain constant. To that end, let $\by_t= (y_{1,t},\ldots,y_{n,t})'$ be an $n \times 1$ vector of endogenous variables at time $t$. The TVP-VAR of \citet{Primiceri05} can be reparameterized in the following structural form:
\begin{equation} \label{eq:SVAR}
	\bA_{t} \by_t = \mathbf{b}_t + \bB_{1,t} \by_{t-1} + \cdots + \bB_{p,t} \by_{t-p} + \vepsilon_t^y, \quad \vepsilon_t^y \sim \distn{N}(\mathbf{0}, \vSigma_t),
\end{equation}
where $\mathbf{b}_t$ is an $n\times 1 $ vector of time-varying intercepts, $\bB_{1,t}, \ldots, \bB_{p,t}$ are $n \times n$ VAR coefficient matrices, $\bA_{t}$ is an $n \times n$ lower triangular matrix with ones on the diagonal and $\vSigma_t = \diag(\exp(h_{1,t}), \ldots, \exp(h_{n,t}))$. The law of motion of the VAR coefficients and log-volatilites will be specified below. Since the system in \eqref{eq:SVAR} is written in the structural form, the covariance matrix $\vSigma_t$ is diagonal by construction. Consequently, we can estimate this recursive system equation by equation without loss of efficiency.

We note that \citet*{CCM19} pioneer a similar equation-by-equation estimation approach for a large reduced-form constant-coefficient VAR with stochastic volatility. The main advantage of the structural-form representation is that it allows us to rewrite the VAR as $n$ unrelated regressions, and it leads to a more efficient sampling scheme. The main drawback of this representation, however, is that the implied reduced-form estimates depend on how the variables are ordered in the system. We will investigate the extent to which these estimates depend on the ordering in Section~\ref{ss:order}.

\subsection{An Equation-by-Equation Representation}

It is convenience to introduce some notations. Let $b_{i,t}$ denote the $i$-th element of $\mathbf{b}_t$ and let $\bB_{i,j,t}$ represent the $i$-th row of $\bB_{j,t}$. Then, $\vbeta_{i,t} = (b_{i,t},\bB_{i,1,t},\ldots,\bB_{i,p,t})'$ is the intercept and VAR coefficients of the $i$-th equation and is of dimension $k_{\beta} \times 1$  with $k_{\beta} = np+1$. Moreover, let $\valpha_{i,t} $ denote the free elements in the $i$-th row of the contemporaneous impact matrix $\bA_t$ for $i=2,\ldots, n$. That is, $\valpha_{i,t} = (A_{i1,t},\ldots, A_{i(i-1),t})'$ is of dimension $k_{\alpha_i}\times 1$ with $k_{\alpha_i} = i-1$. Then, the $i$-th equation of the system in \eqref{eq:SVAR} can be rewritten as:
\[
    y_{i,t} = \tilde{\bx}_t \vbeta_{i,t} + \tilde{\bw}_{i,t}\valpha_{i,t} + \epsilon_{i,t}^y, 
			\quad \epsilon_{i,t}^y \sim \distn{N}(0, \e^{h_{i,t}}),
\]
where $\tilde{\bx}_t = (1, \by_{t-1}',\ldots, \by_{t-p}')$ and $\tilde{\bw}_{i,t} = (-y_{1,t},\ldots, -y_{i-1,t})$. Note that $y_{i,t}$ depends on the contemporaneous variables $y_{1,t},\ldots, y_{i-1,t}$. But since the system is triangular, when we perform the change of variables from $\vepsilon_t^y$ to $\by_t$  to obtain the likelihood function, the density function remains Gaussian.

If we let $\bx_{i,t} = (\tilde{\bx}_t, \tilde{\bw}_{i,t})$, we can further simplify the $i$-th equation as:
\begin{equation} \label{eq:yt}
	y_{i,t} = \bx_{i,t} \vtheta_{i,t} + \epsilon_{i,t}^y, \quad \epsilon_{i,t}^y \sim \distn{N}(0,\e^{h_{i,t}}),
\end{equation}
where $\vtheta_{i,t} = (\vbeta_{i,t}', \valpha_{i,t}')'$ is of dimension $k_{\theta_i} = k_{\beta} + k_{\alpha_i} = np+i.$ Hence, we have rewritten the TVP-VAR in~\eqref{eq:SVAR} as $n$ unrelated regressions. 
Finally, the coefficients and log-volatilities are assumed to evolve as independent random walks:
\begin{align}	
	\vbeta_{i,t} & = \vbeta_{i,t-1} + \vepsilon_{i,t}^{\beta}, & \vepsilon_{i,t}^{\beta} 
		& \sim \distn{N}(\mathbf{0}, \vSigma_{\beta_i}), \label{eq:betat}	\\
	\valpha_{i,t} & = \valpha_{i,t-1} + \vepsilon_{i,t}^{\alpha}, & \vepsilon_{i,t}^{\alpha} 
		& \sim \distn{N}(\mathbf{0}, \vSigma_{\alpha_i}), \label{eq:alphat}	\\
   h_{i,t} & = h_{i,t-1} + \epsilon_{i,t}^h, & \epsilon_{i,t}^h & \sim \distn{N}(0, \sigma_{h,i}^2), 
	\label{eq:ht}
\end{align}
where the initial conditions $\vbeta_{i,0}, \valpha_{i,0}$ and $h_{i,0}$ are treated as unknown parameters to be estimated. The system in \eqref{eq:yt}--\eqref{eq:ht} specifies a reparameterization of a standard TVP-VAR in which all equations have time-varying parameters and stochastic volatility.

Note that the innovations in \eqref{eq:betat}-\eqref{eq:ht} are assumed to be independent across equations. This assumption is partly motivated by the concern of proliferation of correlation parameters, especially when $n$ is large, if the correlations of the innovations are unrestricted. 
In addition, for $\vbeta_{i,t}$ and $\valpha_{i,t}$, this independence assumption is important for extending the setup later so that we can turn on and off the time variation in both equations. In contrast, it is feasible to allow the innovations to $h_{i,t}$ to be correlated across equations (with a slight increase of computational cost). In preliminary work we considered such an extension. While the estimation results suggest that the correlation parameters are sizable, this extension leads to only very modest forecast gains (see Appendix~D for details). Therefore, in what follows we maintain the independence assumption in \eqref{eq:betat}-\eqref{eq:ht} as the baseline. 

\subsection{The Non-Centered Parameterization}

Next, we introduce a framework that allows the model to determine in a data-driven fashion whether the VAR coefficients and the contemporaneous relations among the endogenous variables in each equation are time varying or constant. For that purpose, we adapt the non-centered parameterization of \citet{FSW10} to our hybrid TVP-VARs. 
More specifically, for $i=1,\ldots, n, t=1,\ldots, T,$ we consider the following model:
\begin{align}
	y_{i,t} & = \bx_{i,t} \vtheta_{i,0} + \gamma_i^{\beta}\tilde{\bx}_{t}\vSigma_{\beta_i}^{\frac{1}{2}}\tilde{\vbeta}_{i,t} + \gamma_i^{\alpha}\tilde{\bw}_{i,t}\vSigma_{\alpha_i}^{\frac{1}{2}}\tilde{\valpha}_{i,t} + \epsilon_{i,t}^y, & \epsilon_{i,t}^y & \sim \distn{N}(0,\e^{h_{i,t}}), \label{eq:yi}  \\
	\tilde{\vbeta}_{i,t} & = \tilde{\vbeta}_{i,t-1} + \tilde{\vepsilon}_{i,t}^{\beta}, & \tilde{\vepsilon}_{i,t}^{\beta} & \sim \distn{N}(\mathbf{0}, \mathbf{I}_{k_{\beta}}), \label{eq:betai} 	\\
  \tilde{\valpha}_{i,t} & = \tilde{\valpha}_{i,t-1} + \tilde{\vepsilon}_{i,t}^{\alpha}, & \tilde{\vepsilon}_{i,t}^{\alpha} & \sim \distn{N}(\mathbf{0}, \mathbf{I}_{k_{\alpha_i}}), \label{eq:alphai} \\
	h_{i,t} & = h_{i,t-1} + \epsilon_{i,t}^h, & \epsilon_{i,t}^h & \sim \distn{N}(0, \sigma_{h,i}^2), \label{eq:hi}
\end{align}
where $\tilde{\vbeta}_{i,0} = \mathbf{0}$ and $ \tilde{\valpha}_{i,0} = \mathbf{0}$. Here $\gamma_i^{\beta}$ and  $\gamma_i^{\alpha}$ are indicator variables that take values of either 0 or 1. 

The model in~\eqref{eq:yi}-\eqref{eq:hi} includes a wide variety of popular VAR specifications. For example, assuming that all indicators take the value of 1, the above model is just a reparameterization of the TVP-VAR in \eqref{eq:yt}--\eqref{eq:ht}. To see that, 
define $\vbeta_{i,t} = \vbeta_{i,0} + \gamma_i^{\beta}\vSigma_{\beta_i}^{\frac{1}{2}}\tilde{\vbeta}_{i,t}$ and $\valpha_{i,t} = \valpha_{i,0} + \gamma_i^{\alpha}\vSigma_{\alpha_i}^{\frac{1}{2}}\tilde{\valpha}_{i,t}$. Then, when $\gamma_i^{\beta} = \gamma_i^{\alpha} = 1, i=1,\ldots, n$, it is clear that \eqref{eq:yi} becomes~\eqref{eq:yt}. In addition, we have
\begin{align*}
	\vbeta_{i,t} - \vbeta_{i,t-1} & = \vSigma_{\beta_i}^{\frac{1}{2}}(\tilde{\vbeta}_{i,t} - \tilde{\vbeta}_{i,t-1}) = \vSigma_{\beta_i}^{\frac{1}{2}}\tilde{\vepsilon}_{i,t}^{\beta}, \\	
	\valpha_{i,t} - \valpha_{i,t-1} & = \vSigma_{\alpha_i}^{\frac{1}{2}}(\tilde{\valpha}_{i,t} - \tilde{\valpha}_{i,t-1}) = \vSigma_{\alpha_i}^{\frac{1}{2}}\tilde{\vepsilon}_{i,t}^{\alpha}.
\end{align*}
Hence, $\vbeta_{i,t}$ and $\valpha_{i,t}$ follow the same random walk processes as in \eqref{eq:betat} and \eqref{eq:alphat}, respectively. We have therefore shown that when $\gamma_i^{\beta} = \gamma_i^{\alpha} = 1, i=1,\ldots, n$, the proposed model reduces to a TVP-VAR with stochastic volatility.

For the intermediate case where $\gamma_i^{\beta} = 1$ and $ \gamma_i^{\alpha} = 0, i=1,\ldots, n$, the proposed model reduces to a structural-form reparameterization of the model in \citet{CS05}, i.e., a TVP-VAR with stochastic volatility but the contemporaneous relations among the endogenous variables are restricted to be constant. In the extreme case where $\gamma_i^{\beta} = \gamma_i^{\alpha} = 0, i=1,\ldots, n$, the proposed model then becomes a constant-coefficient VAR with stochastic volatility---a reparameterization of the specification in \citet{CCM19}. More generally, by allowing the indicators $\gamma_i^{\beta}$ and $\gamma_i^{\alpha}$ to take different values, we can have a VAR in which only some equations have time-varying parameters. Note that it is straightforward to include a few additional indicators to allow for more flexible forms of time variation. For example, one can replace $\gamma_i^{\beta}$ with two indicators, say, $\gamma_i^{\beta, \text{own}}$ and  $\gamma_i^{\beta, \text{other}}$, which control the time variation in the elements of $\vbeta_{i,t}$ that correspond to coefficients on own lags and lags of other variables, respectively. The posterior simulator in Section~\ref{ss:estimation} can be modified to handle this case, with a slight increase in computation time.

These indicators are not fixed but are estimated from the data. More precisely, we specify that each $\gamma^\beta_i$ follows an independent Bernoulli distribution with success probability $\Pm(\gamma^\beta_i = 1) = p^{\beta}_i, i=1,\ldots, n$. Similarly for $\gamma^{\alpha}_i$: $\Pm(\gamma^{\alpha}_i = 1) = p^{\alpha}_i$. These success probabilities $p^{\beta}_i$ and $p^{\alpha}_i, i=1,\ldots, n$, are in turn treated as parameters to be estimated. In contrast to typical setups where time variation in parameters is assumed \citep[e.g.][]{CS01,CS05,Primiceri05}, here the proposed model puts positive probabilities in simpler models in which the VAR coefficients and the contemporaneous relations among the variables are constant. The values of the indicators are determined by the data, and these time-varying features are turned on only when they are warranted. The proposed model therefore is not only flexible in the sense that it includes a wide variety of specifications popular in applied work as special cases, it also induces parsimony to combat over-parameterization concerns.

\subsection{An Exploration of the  Model Space } \label{ss:BMA}

The proposed hybrid TVP-VAR can also be viewed as a Bayesian model average of a wide variety TVP-VARs with different forms of time variation. To see that, let $\vgamma = (\vgamma_1,\ldots, \vgamma_n)'$ denote the vector of indicator variables with $\vgamma_i = (\gamma^\beta_i, \gamma^\alpha_i)$. Note that each value of $\vgamma\in\{0,1\}^{2n}$ corresponds to a particular TVP-VAR in which the time variation of the $i$-th equation is characterized by $\vgamma_i$. For example, $\vgamma = \mathbf{0}$ corresponds to a constant-coefficient VAR with stochastic volatility. Then, the posterior distribution of any model parameters under the proposed model can be represented as the posterior average with respect to $p(\vgamma\gvn \by)$, i.e., the posterior model probabilities of the collection of $2^{2n}$ TVP-VARs with different forms of time variation, where $\by$ denotes the data. For example, the joint distribution of $\vbeta$ and $\valpha$, the time-varying VAR coefficients and free elements of the contemporaneous impact matrix, can be represented as
\[
	p(\vbeta,\valpha\gvn\by) = \sum_{\bc\in\{0,1\}^{2n}} p(\vbeta,\valpha \gvn\by,\vgamma=\bc) p(\vgamma=\bc\gvn\by).
\]
For a small VAR with $n=3$ variables (and the additional assumption that $\gamma_i^{\beta} = \gamma_i^{\alpha}$), \citet{CE18b} estimate all $2^3=8$ TVP-VARs and the corresponding posterior model probabilities. For larger $n$, this approach of computing $p(\vgamma=\bc\gvn\by)$ and sampling from $p(\vbeta,\valpha \gvn\by,\vgamma=\bc)$ for all $2^{2n}$ possible models is clearly infeasible. In contrast, by including the model indicator $\vgamma$ in the estimation, we simultaneously explore the parameter space and the model space. This latter approach is convenient and computationally feasible for large systems.

It is also instructive to investigate how the value of the model indicator $\vgamma$ is determined by the data. To fix ideas, suppose we wish to compare two TVP-VARs, represented as $\vgamma=\bc_1$ and $\vgamma=\bc_2$. 
Let $p(\by \gvn\vgamma=\bc_j)$ denote the marginal likelihood under model $\vgamma=\bc_j, j=1,2$, i.e.,
\begin{equation}\label{eq:ML}
	p(\by \gvn\vgamma=\bc_j) = \int p(\by \gvn \vpsi_j, \vgamma=\bc_j) p(\vpsi_j\gvn \vgamma=\bc_j)\di\vpsi_j,
\end{equation}
where $\vpsi_j$ is the collection of model-specific time-invariant parameters and time-varying states (in our setting these parameters and states are common across models and $\vpsi_1=\vpsi_2$), $p(\by \gvn \vpsi_j, \vgamma=\bc_j)$ is the (complete-data) likelihood and $p(\vpsi_j\gvn \vgamma=\bc_j)$ is the prior density. Then, the posterior odds ratio in favor of model $\vgamma=\bc_1$ against model $\vgamma=\bc_2$ is given by:
\[
	\frac{p(\vgamma=\bc_1 \gvn\by)}{p(\vgamma=\bc_2\gvn\by)} = \frac{p(\vgamma=\bc_1)}{p(\vgamma=\bc_2)}\times 
	\frac{p(\by\gvn \vgamma=\bc_1)}{p(\by\gvn \vgamma=\bc_2)},
\]
where $p(\vgamma=\bc_1)/p(\vgamma=\bc_2)$ is the prior odds ratio. It follows that if both models are equally probable \textit{a priori}, i.e., $p(\vgamma=\bc_1) = p(\vgamma=\bc_2)$, the posterior odds ratio between the two models is then equal to the ratio of the two marginal likelihoods, or the Bayes factor. More generally, under the assumption that each TVP-VAR has the same prior probability, the value of the model indicator $\vgamma$ is determined by the marginal likelihood $p(\by\gvn \vgamma)$. That is, if the TVP-VAR represented by $\vgamma=\bc$ forecasts the data better (as one-step-ahead density forecasts), the value $\bc$ will have a higher weight.

\section{Priors and Bayesian Estimation} \label{s:estimation}

In this section we first describe in detail the priors on the time-invariant parameters. We then outline the posterior simulator to estimate the model described in \eqref{eq:yi}--\eqref{eq:hi}

\subsection{Priors} \label{ss:priors}

For notational convenience, stack $\by_{i} = (y_{i,1},\ldots, y_{i,T})'$, $\vbeta_{i} = (\vbeta_{i,1}',\ldots, \vbeta_{i,T}')'$, $\valpha_{i} = (\valpha_{i,1}',\ldots, \valpha_{i,T}')'$ and $\bh_{i} = (h_{i,1},\ldots, h_{i,T})'$ over $t=1,\ldots, T$, and collect $\by = \{\by_{i}\}_{i=1}^n$, $\vbeta = \{\vbeta_{i}\}_{i=1}^n $, $\valpha = \{\valpha_{i}\}_{i=1}^n $ and $\bh=\{\bh_{i}\}_{i=1}^n$ over $i=1,\ldots,n$, and similarly define $\widetilde{\vbeta}_i$ and $\widetilde{\valpha}_i$. Furthermore, let $\vSigma_{\theta_i} = \text{diag}(\vSigma_{\beta_i},\vSigma_{\alpha_i})$ and $\vgamma_i = (\gamma^\beta_i, \gamma^\alpha_i)$. In our model, the time-invariant parameters are $\vgamma = (\vgamma_1,\ldots, \vgamma_n)'$, $\vSigma_\theta=\{\vSigma_{\theta_i}\}_{i=1}^n$, $\vSigma_h = \{\sigma_{i,h}^2\}_{i=1}^n$, $\vtheta_0= (\vtheta_{1,0}',\ldots, \vtheta_{n,0}')'$, $\bh_0 = (h_{1,0},\ldots, h_{n,0})'$, $\mathbf{p}^{\beta}=(p^{\beta}_1,\ldots,p^{\beta}_n)'$ and $\mathbf{p}^{\alpha}=(p^{\alpha}_1,\ldots,p^{\alpha}_n)'$. Below we give the details of the priors on these time-invariant parameters.

Since $\vtheta_0= (\vbeta_0',\valpha_0')'$, the initial conditions of the VAR coefficients, is high-dimensional when $n$ is large, appropriate shrinkage is crucial. We assume a Minnesota-type prior on $\vtheta_0$ along the lines in \citet{SZ98}; see also \citet{DLS84}, \citet{litterman86} and \citet{KK97}.
We refer the readers to \citet{KK10}, \citet{DS12} and \citet{karlsson13} for a textbook discussion of the Minnesota prior. More specifically, consider $\vtheta_0\sim\distn{N}(\mathbf{a}_{\vtheta_0},\bV_{\vtheta_0})$, where the prior mean $\vtheta_0$ is set to be zero when the variables are in growth rate to induce shrinkage and the prior covariance matrix $\bV_{\vtheta_0}$ is block-diagonal with $\bV_{\vtheta_0}=\text{diag}(\bV_{\vtheta_{1,0}},\ldots,\bV_{\vtheta_{n,0}})$---here $\bV_{\vtheta_{i,0}}$ is the prior covariance matrix for $\vtheta_{i,0}, i=1,\ldots, n$. For each $\bV_{\vtheta_{i,0}},$ we in turn assume it to be diagonal with the $k$-th diagonal element 
$(V_{\vtheta_{i,0}})_{kk}$ set to be:
\[
	(V_{\vtheta_{i,0}})_{kk} = \left\{
	\begin{array}{ll}
			\frac{\kappa_1}{l^2}, & \text{for the coefficient on the $l$-th lag of variable } i,\\
			\frac{\kappa_2 s_i^2}{l^2 s_j^2}, & \text{for the coefficient on the $l$-th lag of variable } j, j\neq i, \\
			\frac{\kappa_3 s_i^2}{s_j^2}, & \text{for the $j$-th element of } \valpha_i, \\
			\kappa_4 s_i^2, & \text{for the intercept}, \\
	\end{array} \right.
\]
where $s_r^2$ denotes the sample variance of the residuals from regressing $y_{r,t}$ on $\by_{t-1},\ldots,\by_{t-4}$, $r=1,\ldots, n$. Here the prior covariance matrix $\bV_{\vtheta_0}$ depends on four hyperparameters---$\kappa_1,\kappa_2, \kappa_3$ and $\kappa_4$---that control the degree of shrinkage for different types of coefficients. For simplicity, we set $\kappa_3 = 1$ and $\kappa_4 = 100$. These values imply moderate shrinkage for the coefficients on the contemporaneous variables and no shrinkage for the intercepts. 

The remaining two hyperparameters are $\kappa_1$ and $\kappa_2$, which control the overall shrinkage strength for coefficients on own lags and those on lags of other variables, respectively. Departing from \citet{SZ98}, here we allow $\kappa_1$ and $\kappa_2$ to be different, as one might expect that coefficients on lags of other variables would be on average smaller than those on own lags. In fact, \citet{CCM15} and \citet{chan21} find empirical evidence in support of this so-called cross-variable shrinkage. In addition, we treat $\kappa_1$ and $\kappa_2$ as unknown parameters to be estimated rather than fixing them to some subjective values. This is motivated by a few recent papers, such as \citet{CCM15} and \citet*{GLP15}, which show that by selecting this type of overall shrinkage hyperparameters in a data-based fashion, one can substantially improve the forecast performance of the resulting VAR. In addition, this data-based Minnesota prior is also found to forecast better than many recently introduced adaptive shrinkage priors such as the normal-gamma prior, the Dirichlet-Laplace prior and the horseshoe prior. For example, this is demonstrated in a comprehensive forecasting exercise in \citet{CHP20}.

We assume gamma priors for the hyperparameters $\kappa_1$ and $\kappa_2$: $\kappa_j\sim \distn{G}(c_{1,j},c_{2,j}), j=1,2$. We set $c_{1,1} = c_{1,2} = 1 $, 
$c_{2,1} = 1/0.04$ and $c_{2,2} = 1/0.04^2$. These values imply that the prior modes are at zero, which provides global shrinkage. The prior means of $\kappa_1$ and $\kappa_2$ are 0.04 and $0.04^2$ respectively, which are the fixed values used in \citet*{CCM15}. Next, following \citet{FSW10}, the square roots of the diagonal elements of  $\vSigma_{\theta_i} = \diag(\sigma_{\theta_i, 1}^2 , \ldots, \sigma_{\theta_i, k_{\theta_i}}^2), i=1,\ldots, n,$ are independently distributed as mean 0 normal random variables: $\sigma_{\theta_i, j}\sim \distn{N}(0,S_{\theta_i, j}), i=1,\ldots, n, j = 1,\ldots, k_{\theta_i}$. We assume each $\sigma_{h,i}^2$ follow a conventional inverse-gamma priors: $\sigma_{h, i}^2\sim \distn{IG}(\nu_{h,i},S_{h, i}), i=1,\ldots, n$. The success probabilities $p^{\beta}_i$ and $p^{\alpha}_i$ are assumed to have beta distributions: $p^{\beta}_i \sim\distn{B}(a_{p^\beta},b_{p^\beta})$ and $p^{\alpha}_i\sim\distn{B}(a_{p^{\alpha}},b_{p^{\alpha}}), i=1,\ldots, n$. Finally, the elements of the initial condition $\bh_0$ are assumed to be Gaussian: $h_{i,0}\sim\distn{N}(a_{h_{i,0}}, V_{h_{i,0}})$. 

\subsection{The Posterior Simulator} \label{ss:estimation}

We now turn to the estimation of the model in \eqref{eq:yi}--\eqref{eq:hi} given the prior described in the previous section. There are a few challenges in the estimation. First, since $\vbeta_i$ becomes degenerate when $\gamma^{\beta}_i=0$, making its sampling nonstandard (similarly for $\valpha_i$). To sidestep this problem, we will use the parameterization in terms of $\tilde{\vbeta}_i$ and $\tilde{\valpha}_i$. Then, given the posterior draws of $\tilde{\vbeta}_i$, $\tilde{\valpha}_i$ and other parameters, we can recover the posterior draws of  $\vbeta_i$ and $\valpha_i$ using the definitions $\vbeta_{i,t} = \vbeta_{i,0} + \gamma_i^{\beta}\vSigma_{\beta_i}^{\frac{1}{2}}\tilde{\vbeta}_{i,t}$ and $\valpha_{i,t} = \valpha_{i,0} + \gamma_i^{\alpha}\vSigma_{\alpha_i}^{\frac{1}{2}}\tilde{\valpha}_{i,t}$.

Second, since $\tilde{\vbeta}_i$ and the indicator $\gamma_i^\beta$ enter the likelihood in \eqref{eq:yi} multiplicatively, it is vital to sample them jointly (similarly for $\tilde{\valpha}_i$ and $\gamma_i^{\alpha}$); otherwise the Markov chain might get stuck. To see this, consider a simpler sampling scheme in which we simulate $\tilde{\vbeta}_i$ given $\gamma_i^\beta$, followed by sampling $\gamma_i^\beta$ given $\tilde{\vbeta}_i$. Suppose $\gamma_i^\beta = 0$ in the last iteration. Given $\gamma_i^\beta = 0$, $\tilde{\vbeta}_i$ does not enter the likelihood and we simply sample it from its state equation. Since the sampled $\tilde{\vbeta}_i$ has no relation to the data, the implied time variation in the VAR coefficients would not match the data. Consequently, it is highly likely that the model would prefer no time variation, i.e., $\gamma_i^\beta=0$. Hence, it is unlikely for the Markov chain to move away from  $\gamma_i^\beta = 0$ once it is there. It is therefore necessary to sample both $\tilde{\vbeta}_i$ and $\gamma_i^\beta$ in the same step. In addition, since the pair $(\tilde{\vbeta}_i, \gamma_i^\beta)$ and $(\tilde{\valpha}_i, \gamma_i^\alpha)$ enters the likelihood additively, we sample them jointly to further improve efficiency. 

Next, define $\tilde{\vtheta}_{i} = (\tilde{\vtheta}_{i,1}',\ldots, \tilde{\vtheta}_{i,T}')'$ with $\tilde{\vtheta}_{i,t} = (\tilde{\vbeta}_{i,t}',\tilde{\valpha}_{i,t}')'$. Then, one can simulate from the joint posterior distribution using the following posterior sampler that sequentially samples from:
\begin{enumerate}

	\item $p(\vgamma_i, \tilde{\vtheta}_i \gvn  \by, \bh, \vtheta_0, \bh_0, \vSigma_\theta, 
	\vSigma_h, \mathbf{p}^\beta,\mathbf{p}^\alpha, \vkappa)$, $i=1,\ldots, n$;

	\item $p(\bh_i \gvn  \by, \tilde{\vtheta}, \vtheta_0, \bh_0, \vSigma_\theta, \vSigma_h, 
	\vgamma, \mathbf{p}^\beta, \mathbf{p}^\alpha,\vkappa), i=1,\ldots, n$;	

	\item $p(\vSigma_{\theta_i}^{\frac{1}{2}}, \vtheta_{i,0} \gvn \by, 
	\tilde{\vtheta}, \bh, \bh_0, \vSigma_h,\vgamma,\mathbf{p}^\beta,\mathbf{p}^\alpha,\vkappa)$,$i=1,\ldots, n$;

	\item $p(\sigma_{h,i}^2 \gvn  \by, \tilde{\vtheta}, \bh, \bh_0, \vSigma_{\theta}, \vtheta_{0}, 
	\vgamma, \mathbf{p}^\beta, \mathbf{p}^\alpha,\vkappa)$, $i=1,\ldots, n$;
	
	\item $p(h_{i,0} \gvn  \by, \tilde{\vtheta}, \bh, \vSigma_{\theta}, \vtheta_{0}, 
	\vgamma, \vSigma_h, \mathbf{p}^\beta, \mathbf{p}^\alpha,\vkappa)$, $i=1,\ldots, n$;
		
	\item $p(p^\beta_i,p^\alpha_i \gvn \by, \tilde{\vtheta}, \bh, \vSigma_{\theta}, \vtheta_{0},
	\vgamma,\vSigma_h, \bh_0, \vkappa)$, $i=1,\ldots, n$;
	
	\item $p(\vkappa \gvn \by, \tilde{\vtheta}, \bh, \vSigma_{\theta}, \vtheta_{0}, \vgamma, 
	\vSigma_h, \bh_0, \mathbf{p}^\beta,\mathbf{p}^\alpha)$.
		
\end{enumerate}	

Step 2 to Step 7 mainly involve standard sampling techniques and we leave the details to Appendix~A. Here we focus on the first step. 

\underline{\textbf{Step 1}}. We sample the four blocks of parameters $\tilde{\vbeta}_i, \gamma_i^\beta, \tilde{\valpha}_i$ and $\gamma_i^\alpha$ jointly to improve efficiency. This is done by first drawing the indicators
 $\vgamma_i = (\gamma^\beta_i, \gamma^\alpha_i)$ marginally of $\tilde{\vtheta}_{i,t} = (\tilde{\vbeta}_{i,t}', \tilde{\valpha}_{i,t}')'$---but conditional on other parameters---and then sample $\tilde{\vbeta}_i$ and $\tilde{\valpha}_i$ from their joint conditional distribution. The latter of these two steps is straightforward because given $\gamma^\beta_i$ and $\gamma^\alpha_i$, we have a linear Gaussian state space model in $\vtheta_{i,t}$. Specifically, we stack the observation equation \eqref{eq:yi} over $t=1,\ldots, T$:
\begin{align*}
    \by_i = \bX_i\vtheta_{i,0} + \bZ_{\vgamma_i}\tilde{\vtheta}_i + \vepsilon_i^y, \qquad 
		\distn{N}(\mathbf{0},\vSigma_{\bh_i}),
\end{align*}
where $\vSigma_{\bh_i} = \text{diag}(\e^{h_{i,1}},\ldots, \e^{h_{i,T}})$,
\[
    \bX_i = \begin{pmatrix} \bx_{i,1} \\ \vdots \\ \bx_{i,T} \end{pmatrix}, \quad  
		\bZ_{\vgamma_i} = \begin{pmatrix}
    (\gamma_i^{\beta}\tilde{\bx}_{1},\gamma_i^{\alpha}\tilde{\bw}_{i,1})\vSigma_{\theta_i}^{\frac{1}{2}} & \mathbf{0} & \cdots &  \mathbf{0} \\
    \mathbf{0}  & (\gamma_i^{\beta}\tilde{\bx}_{2},\gamma_i^{\alpha}\tilde{\bw}_{i,2})\vSigma_{\theta_i}^{\frac{1}{2}} & \cdots &  \mathbf{0} \\
    \vdots  & \vdots & \ddots &  \vdots \\
    \mathbf{0} & \mathbf{0} & \cdots & (\gamma_i^{\beta}\tilde{\bx}_{T},\gamma_i^{\alpha}\tilde{\bw}_{i,T})\vSigma_{\theta_i}^{\frac{1}{2}} \end{pmatrix}.
\]
Here note that the matrix $\bZ_{\vgamma_i}$ depends on the indicators 
$\vgamma_i=(\gamma_i^{\beta}, \gamma_i^{\alpha})$. Next, stack the state equations \eqref{eq:betai}-\eqref{eq:alphai} over $t=1,\ldots, T$:
\[
    \bH_{k_{\theta_i}}\tilde{\vtheta}_i = \vepsilon_i^{\tilde{\theta}}, \quad \vepsilon_i^{\tilde{\theta}} \sim \distn{N}(\mathbf{0},\mathbf{I}_{T k_{\theta_i}}),
\]
where $\bH_{k_{\theta_i}}$ is the first difference matrix of dimension $k_{\theta_i}= k_{\beta} + k_{\alpha_i}$. Since $\bH_{k_{\theta_i}}$ is a square matrix with unit determinant, it is invertible.
It then follows that $\tilde{\vtheta}_i \sim \distn{N}(\mathbf{0},(\bH_{k_{\theta_i}}'\bH_{k_{\theta_i}})^{-1}).$ Finally, using standard linear regression results, we have
\begin{equation} \label{eq:thetai_cond}
  (\tilde{\vtheta}_i \gvn  \by_i, \bh_i, \vSigma_{\theta_i}, \vtheta_{i,0},\vgamma_i) \sim \distn{N}\left(\widehat{\widetilde{\vtheta}}_i,\bK_{\widetilde{\vtheta}_i}^{-1}\right),
\end{equation}
where
\begin{equation}\label{eq:thetai_mv}
    \bK_{\widetilde{\vtheta}_i} = \bH_{k_{\theta_i}}'\bH_{k_{\theta_i}} + \bZ_{\vgamma_i}'\vSigma_{\bh_i}^{-1}\bZ_{\vgamma_i}, \qquad
    \widehat{\widetilde{\vtheta}}_i =  \bK_{\widetilde{\vtheta}_i}^{-1} \left(\bZ_{\vgamma_i}'\vSigma_{\bh_i}^{-1}(\by_i-\bX_i\vtheta_{i,0})\right).
\end{equation}
Since the precision matrix $\bK_{\widetilde{\vtheta}_i}$ is a band matrix, one can sample $(\tilde{\vtheta}_i \gvn  \by_i, \bh_i, \vSigma_{\theta_i}, \vtheta_{i,0}, \vgamma_i)$ efficiently using the algorithm in \citet{CJ09}. It is worth noting that one could include an additional step to accept or reject the draw $\tilde{\vtheta}_i$ to ensure stationarity by checking the roots of the characteristic polynomial associated with the implied reduced-form VAR coefficients along the lines suggested in \citet{CS05}.

To sample $\vgamma_i=(\gamma_i^{\beta},\gamma_i^{\alpha})$ marginal of $\tilde{\vtheta}_i$, it suffices to compute the four probabilities that $\vgamma_i = (0,0), \vgamma_i=(0,1), \vgamma_i=(1,0),$ and $\vgamma_i=(1,1)$. To that end, note that
\[
    p(\vgamma_i \gvn  \by_i, \bh_i, \vSigma_{\theta_i}, \vtheta_{i,0}) \propto \left[ \int_{\mathbb{R}^{Tk_{\theta_i}}} p(\by_i\gvn \tilde{\vtheta}_i, \bh_i, \vSigma_{\theta_i}, \vtheta_{i,0},\vgamma_i)
    p(\tilde{\vtheta}_i) \di \tilde{\vtheta}_i \right] p(\vgamma_i),
\]
where both the conditional likelihood $p(\by_i\gvn \tilde{\vtheta}_i, \bh_i, \vSigma_{\theta_i}, \vtheta_{i,0},\vgamma_i)$ and the prior density $p(\tilde{\vtheta}_i)$ are Gaussian. 
It turns out that the above integral  admits an analytical expression. In fact, using a similar derivation in \citet{CG16}, one can show that 
\begin{equation} \label{eq:int_theta}
\begin{split}
    \int_{\mathbb{R}^{Tk_{\theta_i}}} p(\by_i & \gvn \tilde{\vtheta}_i, \bh_i, \vSigma_{\theta_i}, \vtheta_{i,0}, \vgamma_i) p(\tilde{\vtheta}_i) \di \tilde{\vtheta}_i \\
    & =  (2\pi)^{-\frac{T}{2}} |\bK_{\tilde{\vtheta}_i}|^{-\frac{1}{2}} \e^{-\frac{1}{2}\left(\sum_{t=1}^T h_{i,t} + (\by_i-\bX_i\vtheta_{i,0})'\vSigma_{\bh_i}^{-1}(\by_i-\bX_i\vtheta_{i,0}) -\widehat{\widetilde{\vtheta}}_i'\bK_{\widetilde{\vtheta}_i}\widehat{\widetilde{\vtheta}}_i\right)},
\end{split}
\end{equation}
where $\widehat{\widetilde{\vtheta}}_i$ and $\bK_{\widetilde{\vtheta}_i}$ are defined in \eqref{eq:thetai_mv}.
Then, one can compute the relevant probabilities using the expression in \eqref{eq:int_theta}. For example, when $\vgamma_i = (0,0)$, $|\bK_{\widetilde{\vtheta}_i}|=1$ and $\widehat{\widetilde{\vtheta}}_i = \mathbf{0}$. It follows that
\begin{equation*} 
\begin{split}
    \Pm(\vgamma_i = (0,0) & \gvn  \by_i, \bh_i, \vSigma_{\theta_i}, \vtheta_{i,0}) \\
		& \propto  (1-p_i^\beta)(1-p_i^\alpha)(2\pi)^{-\frac{T}{2}} \e^{-\frac{1}{2}\left(\sum_{t=1}^T h_{i,t}  + (\by_i-\bX_i\vtheta_{i,0})'\vSigma_{\bh_i}^{-1}(\by_i-\bX_i\vtheta_{i,0})\right)}.
\end{split}
\end{equation*}
Similarly, we have
\begin{equation*} 
\begin{split}
    \Pm(\vgamma_i & = (1,1) \gvn  \by_i, \bh_i, \vSigma_{\theta_i}, \vtheta_{i,0}) \\
    & \propto p_i^{\beta}p_i^{\alpha}(2\pi)^{-\frac{T}{2}}|\bK_{\tilde{\vtheta}_i}(1,1)|^{-\frac{1}{2}}
     \e^{-\frac{1}{2}\left(\sum_{t=1}^T h_{i,t}  + (\by_i-\bX_i\vtheta_{i,0})'\vSigma_{\bh_i}^{-1}(\by_i-\bX_i\vtheta_{i,0}) -\widehat{\widetilde{\vtheta}}_i(1,1)'\bK_{\widetilde{\vtheta}_i}(1,1)\widehat{\widetilde{\vtheta}}_i(1,1)\right)},
\end{split}
\end{equation*}
where $\widehat{\widetilde{\vtheta}}_i(1,1) $ and $\bK_{\widetilde{\vtheta}_i}(1,1)$ denote respectively 
$\widehat{\widetilde{\vtheta}}_i$ and $\bK_{\widetilde{\vtheta}_i}$ evaluated at $\vgamma_i=(1,1)$. The probabilities that $\vgamma_i = (0,1)$ and $\vgamma_i = (1,0)$ can be computed similarly. A draw from this 4-point  distribution is standard once we normalize the probabilities. The details of the remaining steps are provided in Appendix~A.

\section{A Monte Carlo Study} \label{s:MC}

In this section we first conduct a series of simulated experiments to assess how well the posterior sampler works in recovering the time-varying structure in the data generating process. We then document the runtimes of estimating the hybrid TVP-VARs of different dimensions to assess how well the posterior sampler scales to larger systems.

First, we generate 300 datasets from the hybrid VAR in \eqref{eq:yi}--\eqref{eq:hi} with $n=12$ variables and sample size $T=200$, $T=400$ or $T=800$. We set the vector of indicators $\vgamma$ by repeating the four combinations $(0,0), (0,1), (1,0), (1,1)$ three times --- that allows us to study the effect of different combinations of time-varying pattens as well as their positions in the system. We generate $\vbeta_0$, the initial conditions of the VAR coefficients, stochastically as follows. The intercepts are drawn independently from the uniform distribution on the interval $(-10,10)$, i.e., $\distn{U}(-10, 10)$. For the VAR coefficients, the diagonal elements of the first VAR coefficient matrix are iid $\distn{U}(0,0.5)$ and the off-diagonal elements are from $\distn{U}(-0.2,0.2)$. All other elements of the $j$-th ($j > 1$) VAR coefficient matrices are iid $\distn{N}(0,0.1^2/j^2).$ Finally, the elements of $\valpha_0$ are drawn independently from $\distn{U}(-0.5, 0.5)$. 

If the coefficient $\theta_{ij,t}$ is  time-varying (i.e., the associated indicator $\gamma_i^{\alpha}$ or $\gamma_i^{\beta}$ is 1), it is generated from the state equation \eqref{eq:betat} or \eqref{eq:alphat} with $\sigma_{\theta_{i},j}^2 = 0.01^2$ if $\theta_{ij,t}$ is a VAR coefficient and $\sigma_{\theta_{i},j}^2 = 0.1^2$ if it is an intercept for $i=1,\ldots,n, j=1,\ldots, k_{\theta_i}$. Finally, for the log-volatility processes, we draw $h_{0,i}\sim \distn{U}(-2,2)$ and set $\sigma_h^2 = 0.1, i=1,\ldots,n$. 

In the Monte Carlo study we use the priors described in Section~\ref{ss:priors} with the following hyperparameters. The prior means of the initial conditions $\vtheta_0$ and $\bh_0$ are set to be zero 
$\mathbf{a}_{\vtheta_0} = \mathbf{0}$ and $\ba_{h}=\mathbf{0}$, and the prior covariance matrix of $\bh_0$ is $\bV_{h} = 10\times\mathbf{I}_n$. The hyperparameter of $\sigma_{\theta_i,j}$ is set so that the implied prior mean of $\sigma_{\theta_i,j}$ is $0.01^2$ if it is associated with a VAR coefficient and $0.1^2$ for an intercept. Finally, we set the hyperparameters of $\gamma_i^{\beta}$ and $\gamma_i^{\alpha}$ to be $a_{p^\beta} = b_{p^\beta} = a_{p^\alpha} = b_{p^\alpha} = 0.5$. These values imply that the prior modes are at 0 and 1, whereas the prior mean is 0.5.

Given a dataset and the priors described above, we estimate the hybrid VAR using the posterior sampler in Section~\ref{ss:estimation} and obtain the posterior mode of $\vgamma$. We repeat this procedure for all the datasets and compute the frequencies of $\gamma^\beta_i$ and $\gamma^\alpha_i$ being one, $i=1,\ldots, n$. The results are reported in Table~\ref{tab:gam}. 

Overall, the posterior sampler works well and is able to recover the true time-varying structure in the simulated data on average. While it is harder to pin down the correct value of $\gamma^\alpha_i$ compared to $\gamma^\beta_i$, the frequencies of identifying the true value of $\gamma^\alpha_i$ are still reasonably good, even for a small sample of $T=200$. In addition, these results substantially improve when the sample size increases from $T=200$ to $T=800$. All in all, these Monte Carlo results confirm that the proposed hybrid model can recover salient patterns---such as time-varying conditional means and covariances---in the data. 

To further investigate the effect of the beta prior on $\gamma_i^{\beta}$ and $\gamma_i^{\alpha}$, we repeat the Monte Carlo experiments but assume a uniform prior on the unit interval $(0,1)$, i.e., $\gamma_i^{\beta}, \gamma_i^{\alpha}\sim \distn{B}(1,1) = \distn{U}(0,1).$ Hence, both the prior means and modes are 0.5. The Monte Carlo results are similar to the baseline case and they are reported in Appendix~D.

\begin{table}[H]
\caption{Frequencies (\%) of the posterior modes of $\gamma^\beta_i$ and $\gamma^\alpha_i$ being one in 300 datasets.}
\label{tab:gam}
\centering
\begin{tabular}{ccccccccc}\hline\hline
Equation	&	True $\gamma_i^{\beta}$	&	True $\gamma_i^{\alpha}$	&	\multicolumn{2}{c}{$T=200$} &\multicolumn{2}{c}{$T=400$}			&	\multicolumn{2}{c}{$T=800$}			\\
	&		&		&	$\gamma_i^{\beta}$ &	$\gamma_i^{\alpha}$ &	$\gamma_i^{\beta}$ &	$\gamma_i^{\alpha}$ & 
$\gamma_i^{\beta}$	&	$\gamma_i^{\alpha}$	\\ \hline
1	&	0	&	0	&	0.05	&	--	&	0.06	&	--	&	0.05	&	--	\\
\rowcolor{lightgray}
2	&	0	&	1	&	0.02	&	0.73	&	0.04	&	0.88	&	0.05	&	0.93	\\
3	&	1	&	0	&	0.92	&	0.40	&	0.98	&	0.25	&	1.00	&	0.12	\\
\rowcolor{lightgray}
4	&	1	&	1	&	0.94	&	0.58	&	0.98	&	0.64	&	1.00	&	0.75	\\
5	&	0	&	0	&	0.01	&	0.08	&	0.02	&	0.02	&	0.02	&	0.00	\\
\rowcolor{lightgray}
6	&	0	&	1	&	0.01	&	0.93	&	0.03	&	0.96	&	0.04	&	0.98	\\
7	&	1	&	0	&	0.92	&	0.34	&	0.97	&	0.13	&	1.00	&	0.03	\\
\rowcolor{lightgray}
8	&	1	&	1	&	0.91	&	0.68	&	0.95	&	0.80	&	1.00	&	0.94	\\
9	&	0	&	0	&	0.01	&	0.03	&	0.03	&	0.00	&	0.02	&	0.01	\\
\rowcolor{lightgray}
10	&	0	&	1	&	0.06	&	0.90	&	0.04	&	0.94	&	0.10	&	0.99	\\
11	&	1	&	0	&	0.89	&	0.29	&	0.94	&	0.11	&	1.00	&	0.02	\\
\rowcolor{lightgray}
12	&	1	&	1	&	0.92	&	0.77	&	0.93	&	0.88	&	0.99	&	0.96	\\ \hline \hline
\end{tabular}
\end{table}

Next, we document the runtimes of estimating the hybrid TVP-VARs of different sizes to assess how well the posterior sampler scales to higher dimensions. More specifically, Table~\ref{tab:times} reports the runtimes (in minutes) to obtain 1,000 posterior draws from the hybrid models with $n= 10, 20, 30$ variables and $T=400, 800$ time periods. The posterior sampler is implemented in $\matlab$ on a standard desktop with an Intel Core i7-7700 @3.60 GHz processor and 64 GB memory. As a comparison, we also include the corresponding runtimes of fitting the TVP-VAR of \citet{Primiceri05} using the algorithm in \citet{DP15}. Note that the algorithm in \citet{DP15} samples all the time-varying VAR coefficients $\vbeta$ in one block and it tends to be very computationally intensive for larger systems. One potential solution is to develop an equation-by-equation estimation procedure similar to that in \citet{CCCM21}. Since the algorithm is designed for models with a constant contemporaneous impact matrix, extending it to handle the TVP-VAR of \citet{Primiceri05}---which features a time-varying contemporaneous impact matrix---would be an interesting future research direction.

\begin{table}[H]
\centering
\caption{The runtimes (in minutes) to obtain 1,000 posterior draws from the hybrid TVP-VAR with $n$ variables and $T$ time periods. All VARs have $p = 2$ lags.}
\label{tab:times}
\begin{tabular}{lcccccc}
\hline \hline
 	&	\multicolumn{3}{c}{$T = 400$}					&	\multicolumn{3}{c}{$T = 800$} \\			
	&	$n = 10$  	&	$n=20$  	&	$n=30$  	&	$n = 10$  	&	$n=20$  	&	$n=30$ \\ \hline
Hybrid TVP-VAR	&	4	&	29	&	94	&	8	&	59	&	188	\\
\rowcolor{lightgray}
\citet{Primiceri05}	&	12	&	209	&	--	&	25	&	415	&	--	\\ \hline\hline
\end{tabular}
\end{table}

It is evident from the table that for typical applications with 15-30 variables, the proposed model can be estimated reasonably quickly. In addition, using the recursive representation that admits straightforward equation-by-equation estimation, fitting the proposed model is much faster than estimating the TVP-VAR of \citet{Primiceri05}, even though the former is more flexible.
 
\section{Application: Model Comparison and Forecasting} \label{s:application}

In this section we fit a large US macroeconomic dataset set to demonstrate the usefulness of the proposed model. After describing the dataset in Section \ref{ss:data}, we first investigate how different variable orderings affect the estimates from the proposed hybrid TVP-VAR relative to the TVP-VAR of \citet{Primiceri05} in Section~\ref{ss:order}. We then present the full sample results in Section~\ref{ss:fullsample}. In particular, we conduct a formal Bayesian model comparison exercise to shed light on the time-varying patterns of the model parameters. We then consider a pseudo out-of-sample forecasting exercise in Section~\ref{ss:forecasting}. We show that the forecast performance of the proposed model compares favorably to a range of standard benchmarks.

\subsection{Data and Prior Hyperparameters} \label{ss:data}

The US dataset for our empirical application consists of 20 quarterly variables with a sample period from 1959Q1 to 2018Q4. It is sourced from the FRED-QD database at the Federal Reserve Bank of St. Louis as described in \citet{MN21}. Our dataset contains a variety of standard macroeconomic and financial variables, such as Real GDP, industrial production, inflation rates, labor market variables, money supply and interest rates. They are transformed to stationarity, typically to annualized growth rates. The complete list of variables and how they are transformed is given in Appendix~C. 

We use the priors described in Section~\ref{ss:priors}. In particular, since the data are transformed to growth rates, we set the prior mean of $\vtheta_0$ to be zero, i.e., $\mathbf{a}_{\vtheta_0} = \mathbf{0}$. For the prior hyperparameters on $\kappa_1$ and $\kappa_2$, we set $c_{1,1} = c_{1,2} = 1 $, $c_{2,1} = 1/0.04$ and $c_{2,2} = 1/0.04^2$. These values imply that the prior means of $\kappa_1$ and $\kappa_2$ are respectively 0.04 and $0.04^2$. For the hyperparameters of the initial conditions $\bh_0$, we set $\ba_{h}=\mathbf{0}$ and $\bV_{h} = 10\times\mathbf{I}_n$. Next, the hyperparameters of $\sigma_{h,i}^2$ are set so that the prior mean is $0.1$. 
Similarly, the hyperparameters of $\sigma_{\theta_i,j}$ are chosen so that the implied prior mean is $0.01^2$ if it is associated with a VAR coefficient and $0.1^2$ for an intercept. Finally, we set $a_{p^\theta} = b_{p^\theta} = a_{p^h} = b_{p^h} = 0.5$. These values imply prior modes at 0 and 1, whereas the prior mean is 0.5.

\subsection{The Role of Variable Ordering} \label{ss:order}

Since the proposed hybrid TVP-VAR is written in the recursive structural form---which is used as a computational device and not as an identification scheme---one naturally wonders how the assumed order of the variables affects the model estimates compared to standard TVP-VARs such as the model in \citet{Primiceri05}. Conceptually, the proposed hybrid TVP-VAR is not order invariant due to two components. First, the VAR coefficients are in structural form, and their priors induce priors on the reduced-form parameters that depend on the order of the variables. Second, the multivariate stochastic volatility specification is constructed based on the lower triangular impact matrix $\bA_t$. And since priors are independently elicited for $\bA_t$ and the stochastic volatility, the implied prior on the covariance matrix is not order invariant. 

Popular TVP-VARs such as \citet{CS05} and \citet{Primiceri05} share the second component but not the first (since they are formulated as reduced-form VARs); see, e.g., the discussion in \citet{Primiceri05} and \citet{CCM19}. Hence, one might expect that estimates from the proposed hybrid TVP-VAR would be more sensitive to how the variables are ordered compared to those of \citet{Primiceri05}. On the other hand, as discussed in Section~\ref{ss:BMA}, the proposed hybrid TVP-VAR can be viewed as a Bayesian model average of a wide variety of VARs with many different forms of time variation. Since some of these VARs are more parsimonious and have restricted time variation---including the model of \citet{CCM19} that is less prone to the ordering issue---the resulting Bayesian model average estimates could in principle be more robust to different orderings compared to \citet{Primiceri05}. Hence, whether the ordering issue is more severe in the proposed hybrid TVP-VAR relative to \citet{Primiceri05} is an empirical question, and we investigate this issue below.

It is worth noting that we choose the TVP-VAR of \citet{Primiceri05} to be our benchmark, even though it is not order invariant, because it is generally viewed as the state-of-the-art and it is widely used as a reduced-form VAR for both forecasting \citep[e.g.,][]{DGG13} and structural analysis using non-recursive identification schemes \citep[e.g.,][]{benati2008, BP13}. There are a few recent papers that aim to develop order-invariant VARs with multivariate stochastic volatility, such as \citet{Bognanni18}, \citet{SZ20}, \citet{ARRS21} and \citet{CKY21}. These models, however, are either designed for small TVP-VARs or they do not feature time-varying VAR coefficients.

\begin{figure}[H]
    \centering
   \includegraphics[width=.8\textwidth]{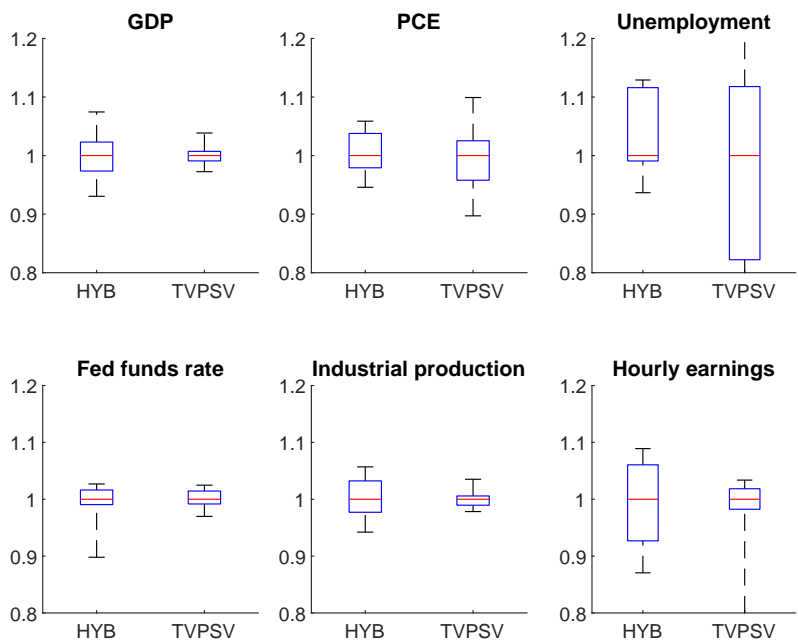}
   \caption{Boxplots of the relative mean squared errors of the fitted means for the proposed hybrid TVP-VAR (HYB) and the model of \citet{Primiceri05} (TVPSV).}
   \label{fig:ordering_yhat}	
\end{figure}

Now, we investigate how different variable orderings impact the estimates from the proposed hybrid TVP-VAR relative to the TVP-VAR of \citet{Primiceri05}. To that end, we use 6 variables---real GDP, PCE inflation, unemployment, Fed funds rate and industrial production and real average hourly earnings in manufacturing---and we consider all $6!=720$ possible orderings. For each ordering of these 6 variables, we fit the proposed model and obtain the fitted values (i.e., the time-varying conditional means). We then compute the mean squared errors (against the observed values) of the 6 variables. We repeat this exercise for the TVP-VAR of \citet{Primiceri05}. Since there are 720 MSEs for each variable and each model, to better summarize the results we report the boxplots of the MSEs in Figure~\ref{fig:ordering_yhat}. The middle line of each box denotes the median, while the lower and upper lines represent, respectively, the 25- and the 75-percentiles. The whiskers extend to the maximum and minimum.

Since the goal of this exercise is to compare the variability of the fitted means, we normalize the MSEs by the medians (so that the red line of each boxplot is one). Overall, the variability of the estimates from the proposed hybrid TVP-VAR is comparable to that of the model of \citet{Primiceri05}. It is also interesting to note that for the majority of the variables the variability is relatively small. One exception is the unemployment rate---this is partly due to the very small base rate (e.g., the MSE of the unemployment is less than 1\% of the MSE of the real GDP).

\begin{figure}[H]
    \centering
   \includegraphics[width=.8\textwidth]{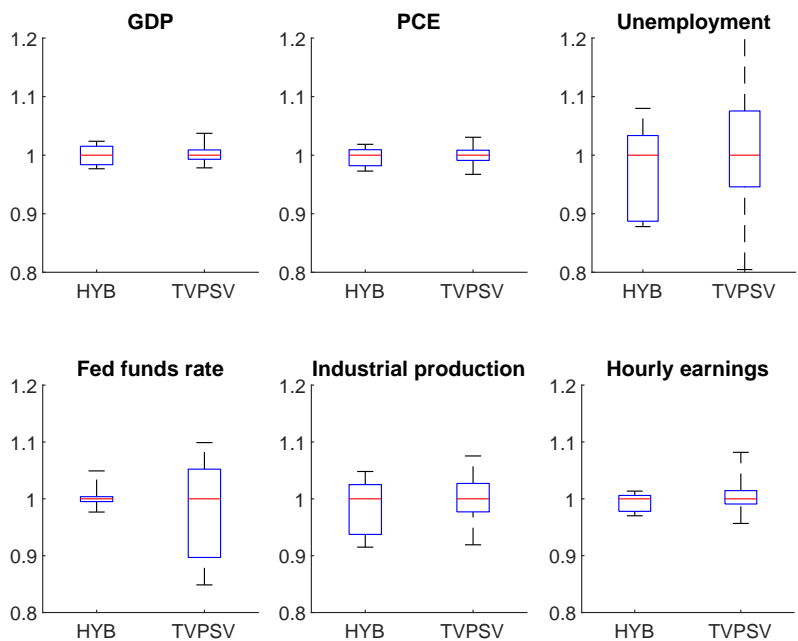}
   \caption{Boxplots of the relative mean squared errors of the estimated variances for the proposed hybrid TVP-VAR (HYB) and the model of \citet{Primiceri05} (TVPSV).}
	\label{fig:ordering_RV}	
\end{figure}

Next, we investigate the variability of the variance estimates due to different orderings. For that purpose, we obtain the fitted variance of each variable in each ordering, and compute the mean squared error against the `realized volatility' values 
(defined as $\text{RV}_{i,t} = (y_{i,t} - \hat{y}_{i,t})^2,$ where $\hat{y}_{i,t}$ is the fitted value from the regression of $y_{i,t}$ on an intercept and $y_{i,t-1}, \ldots, y_{i,t-4}$). The results are reported in Figure~\ref{fig:ordering_RV}. Again, the results show that the variability of the variance estimates from the proposed model is comparable to that of the TVP-VAR of \citet{Primiceri05}, even though the former uses a recursive structural-form representation. 

We also investigate the variability of point and density forecasts similar to the exercise in \citet{ARRS21}. More specifically, for each variable ordering, we compute the root mean squared forecast error and the average of log predictive likelihoods (see Section 5.4 for more details) for each of the 6 variables from the proposed model. Consistent with the results in \citet{ARRS21}, the variability of point forecasts is relatively small. In addition, the variability of the forecasts from the proposed model is similar to a full-fledged TVP-VAR where all the VAR coefficients and elements of the impact matrix are time varying. The details are reported in Appendix~D.

\subsection{Full Sample Results} \label{ss:fullsample}

In this section we report the full sample results of the hybrid TVP-VAR fitted using all $n=20$ variables, which are ordered as listed in Table~\ref{tab:gam_n20}. Of particular interest are the posterior means of $\gamma_i^{\beta}$ and $\gamma_i^{\alpha}$, the indicators that control the time variation in the VAR coefficients and the elements of the impact matrix, respectively. These estimates are reported in Table~\ref{tab:gam_n20}. The results clearly show that while many estimates are essentially 0, others are close to 1. In other words, while the time variation in many equations is essentially turned off, there is strong evidence for time-varying parameters in some equations. Since there is substantial heterogeneity in the time-variation pattern across equations, conventional approaches of either assuming time variation in all equations or restricting all parameters to be constant are unlikely to fit the time-variation pattern well.

\begin{table}[H]
\centering
\caption{Posterior means of $\gamma_i^{\beta}$	and $\gamma_i^{\alpha}$ for the hybrid TVP-VAR with $n=20$ variables.} \label{tab:gam_n20}
\begin{tabular}{llcc}
\hline\hline
Equation	& Mnemonic & $\gamma_i^\beta$	&	$\gamma_i^{\alpha}$	\\ \hline
Real GDP	&	GDPC1	&	0.97	&	--	\\
\rowcolor{lightgray}
PCE inflation	&	PCECTPI	&	0.72	&	0.41	\\
Unemployment	&	UNRATE	&	0	&	0.33	\\
\rowcolor{lightgray}
Fed funds rate	&	FEDFUNDS	&	0	&	1	\\
Industrial production index	&	INDPRO	&	1	&	0.46	\\
\rowcolor{lightgray}
Real average hourly earnings in manufacturing 	&	CES3000000008x	&	0.02	&	1	\\
M1	&	M1REAL	&	0.99	&	0.89	\\
\rowcolor{lightgray}
Real PCE	&	PCECC96	&	0.01	&	0.97	\\
Real disposable personal income	&	DPIC96	&	0	&	0.49	\\
\rowcolor{lightgray}
Industrial production: final products	&	IPFINAL	&	0	&	0.24	\\
All employees: total nonfarm	&	PAYEMS	&	0	&	0.07	\\
\rowcolor{lightgray}
Civilian employment 	&	CE16OV	&	0	&	0.95	\\
Nonfarm business section: hours of all persons	&	HOANBS	&	0	&	0.35	\\
\rowcolor{lightgray}
GDP deflator	&	GDPCTPI	&	0	&	0	\\
CPI 	&	CPIAUCSL	&	0	&	1	\\
\rowcolor{lightgray}
PPI	&	PPIACO	&	0.47	&	1	\\
Nonfarm business sector: real compensation per hour	&	COMPRNFB	&	0	&	1	\\
\rowcolor{lightgray}
Nonfarm business section: real output per hour	&	OPHNFB	&	0	&	0	\\
10-year treasury constant maturity rate 	&	GS10	&	0	&	0	\\
\rowcolor{lightgray}
M2	&	M2REAL	&	1	&	0.94	\\
\hline\hline
\end{tabular}
\end{table}
 
In addition, the estimates of $\gamma_i^{\beta}$ and $\gamma_i^{\alpha}$ for the same equation are often of very different magnitudes, suggesting that time variation in one group of parameters does not necessarily imply time variation in the other. These results thus confirm the usefulness of having two separate indicators for each equation. Overall, while we find evidence for time variation in VAR coefficients and elements of the impact matrix in some equations, not all equations need both forms of time variation. These results therefore highlight the empirical relevance of the proposed hybrid TVP-VAR. 

To assess the efficiency of the proposed posterior sampler, we compute the inefficiency factors of the
posterior draws, which are reported in Appendix D. The values of the inefficiency factors are comparable to those of conventional TVP-VARs. The results thus show that the posterior sampler is efficient in terms of producing posterior draws that are not highly autocorrelated.

Next, we consider a Bayesian model comparison exercise to compare the proposed hybrid TVP-VAR to various VARs with different time-variation patterns. In general, to compare models using the Bayes factor, one would require the computation of the marginal likelihood given in \eqref{eq:ML}. Despite recent advances, obtaining the marginal likelihood for high-dimensional models with multiple latent states remains a nontrivial task. Fortunately, one much simpler approach is available when one wishes to compare nested models. More specifically, for nested models, the Bayes factor can be calculated using the Savage-Dickey density ratio \citep{VW95}, which requires only the estimation of the unrestricted model.  More importantly, no explicit computation of the marginal likelihood is needed. This approach has been used to compute the Bayes factor in many empirical applications, including \citet{KP99}, \citet{DS09} and \citet*{KLS10}.

More specifically, suppose we wish to compare the proposed hybrid TVP-VAR against a TVP-VAR characterized by the vector of indicators $\vgamma = \bc\in\{0,1\}^{2n}$ (e.g., $\bc = \mathbf{0}$ represents a constant-coefficient VAR with stochastic volatility.) Since the latter is a restricted version of the former, the Bayes factor in favor of the unrestricted model can be obtained using the Savage-Dickey density ratio via 
\[
	\text{BF}_{\text{u},\bc} = \frac{p(\vgamma=\bc)}{p(\vgamma=\bc\gvn \by)},
\]
where the numerator and the denominator are, respectively, the marginal prior and posterior densities of $\vgamma$ evaluated at $\bc$. Hence, to compute the relevant Bayes factor, one only needs to evaluate two densities at a point. Intuitively, if $\vgamma =\bc$ is less likely under the posterior density relative to the prior density, i.e., $p(\vgamma=\bc\gvn \by) < p(\vgamma=\bc)$, then it is viewed as evidence against the restriction $\vgamma =\bc$ and the unrestricted model is favored with $\log\text{BF}_{\text{u},\bc} > 0$. Any two nested models can be compared similarly. We provide the technical details on evaluating the two densities at $\vgamma =\bc$ in Appendix B. 

We compare the proposed hybrid TVP-VAR to four VARs with stochastic volatility. For each of these VARs, $\gamma_1^{\beta}= \cdots = \gamma_n^{\beta}$ and $\gamma_1^{\alpha}=\cdots = \gamma_n^{\alpha}$, and they are denoted as HYB-$(\gamma_i^{\beta},\gamma_i^{\alpha})$. The first is a full-fledged TVP-VAR where all the VAR coefficients and elements of the impact matrix are time varying with $\gamma_i^{\beta} = \gamma_i^{\alpha} = 1, i=1,\ldots, n$; this is a structural-form version of the TVP-VAR in \citet{Primiceri05} and we denote this model as HYB-$(1,1)$. The second is a VAR with time-varying VAR coefficients but a constant impact matrix, i.e., $\gamma_i^{\beta} = 1, \gamma_i^{\alpha} = 0, i=1,\ldots, n$, which we denote as HYB-$(1,0)$; this is a structural-form parameterization of the TVP-VAR in \citet{CS05}. The third is a constant-coefficient VAR with $\gamma_i^{\beta} = \gamma_i^{\alpha} = 0, i=1,\ldots, n$, which we denote as HYB-$(0,0)$; this is a structural-form variant of the VAR in \citet{CCM19}. Lastly, we also consider a version in which the VAR coefficients are constant but the elements of the impact matrix are time varying, i.e., $\gamma_i^{\beta} = 0, \gamma_i^{\alpha} = 1, i=1,\ldots, n$, which we denote as HYB-$(0,1)$.

Table~\ref{tab:lBF} reports the log Bayes factors of the proposed hybrid TVP-VAR against these four VARs with stochastic volatility. It is clear that there is overwhelming support for the hybrid TVP-VAR relative to the alternatives. For instance, for $n=20$ the Bayes factor in favor of the hybrid TVP-VAR against the best alternative HYB-$(0,1)$ is $\e^{401} \approx 1.42\times 10^{174}$, suggesting the posterior model probability of the hybrid TVP-VAR is practically 1. These model comparison results are consistent with the estimates of $\vgamma$ reported in Table~\ref{tab:gam_n20}, which indicate substantial heterogeneity in the time-variation pattern across equations---hence, fixing these indicators to either 0 or 1 for all equations is likely to be too restrictive.

\begin{table}[H]
\centering
\caption{Log Bayes factors of the proposed hybrid TVP-VAR against four VARs with stochastic volatility: structural-form versions of \citet{Primiceri05} (HYB-$(1,1)$), \citet{CS05} (HYB-$(1,0)$) and \citet{CCM19} (HYB-$(0,0)$), as well as a VAR with constant VAR coefficients but a time-varying impact matrix (HYB-$(0,1)$). A positive log Bayes factor indicates that the hybrid TVP-VAR is favored.} 
\label{tab:lBF}
\begin{tabular}{lccc}
\hline\hline
	&	$n = 3$  	&	$n=6$  	&	$n=20$  	\\	\hline
HYB-(1,1)	&	18	&	60	&	1003	\\
\rowcolor{lightgray}
HYB-(1,0)	&	12	&	89	&	1035	\\
HYB-(0,1)	&	10	&	54	&	401	\\
\rowcolor{lightgray}
HYB-(0,0)	&	8	&	181	&	2540	\\
\hline \hline
\end{tabular}
\end{table}

To investigate the support for the hybrid TVP-VAR across model dimensions, we also consider a small system ($n=3$) with only real GDP, PCE inflation and unemployment, as well as a medium system ($n=6$) with three additional variables: Fed funds rate, industrial production and real average hourly earnings in manufacturing. For both model dimensions, the hybrid TVP-VAR is strongly favored by the data compared to the alternatives.

It is interesting to note that the best model among the four benchmarks changes across model dimensions. More specifically, for the small system with $n=3$ variables, the best model is HYB-$(0,0)$, the constant-coefficient VAR with stochastic volatility, even though it is the most restrictive among the four VARs. This suggests that the additional flexibility in allowing time-varying coefficients does not sufficiently fit the data better to justify the added model complexity. This is inline with the results in \citet{CE18}, who find that, for fitting a 3-variable US dataset, a constant-coefficient VAR with stochastic volatility performs well relative to various VARs with different forms of time variation according to the marginal likelihood.

While HYB-$(0,0)$ is the preferred model for the small system, when more variables are included HYB-$(0,1)$ is strongly favored. For instance, for $n=20$, allowing for time variation in the impact matrix increases the log marginal likelihood by 2,139---comparing HYB-$(0,0)$ and HYB-$(0,1)$. But further allowing for time variation in the VAR coefficients reduces the log marginal likelihood by 602---comparing HYB-$(0,1)$ and HYB-$(1,1)$. Taken together, these results suggest that it is generally useful to allow for time variation in the impact matrix. In contrast, adding time variation in the VAR coefficients should be done judiciously in large systems: while simply allowing all VAR coefficients to be time-varying can be detrimental, using the proposed data-based approach to add time variation equation-wise can substantially improve model-fit.

\subsection{Forecasting Results} \label{ss:forecasting}

Next, we evaluate the forecast performance of the proposed hybrid TVP-VAR relative to a few standard benchmarks. In particular, we consider 1) a conventional homoscedastic and constant-coefficient VAR; 2) a constant-coefficient VAR with stochastic volatility and a constant impact matrix (by setting all $\gamma_i^{\beta}$ and $\gamma_i^{\alpha}$ to 0; denoted as HYB-$(0,0)$); and 3) a full-fledged TVP-VAR (by setting all the indicators to 1; denoted as HYB-$(1,1)$). All models use all $n=20$ variables. The sample period is from 1959Q1 to 2018Q4, and the evaluation period starts at 1985Q1 and runs till the end of the sample. 

We perform a recursive forecasting exercise using an expanding window. More specifically, in each forecasting iteration $t$, we use only data up to time $t$, denoted as $\by_{1:t}$, to estimate the models. We then evaluate both point and density forecasts. We use the conditional expectation $\Em(y_{i,t+m}\gvn \by_{1:t})$ as the $m$-step-ahead point forecast for variable $i$ and the predictive density $p(y_{i,t+m} \gvn \by_{1:t})$ as the corresponding density forecast. 

The metric used to evaluate the point forecasts from model $M$ is the root mean squared forecast error (RMSFE) defined as
\[
 \text{RMSFE}_{i,m}^M = \sqrt{\frac{\sum_{t=t_0}^{T-m}( y_{i,t+m}^{\text{o}} - \Em(y_{i,t+m}\gvn \by_{1:t}))^2}{T-m-t_0+1}},
\]
where $y_{i,t+m}^{\text{o}}$ is the actual observed value of $y_{i,t+m}$. For RMSFE, a smaller value indicates better forecast performance. To evaluate the density forecasts, the metric we use is the average of log predictive likelihoods (ALPL):
\[
  \text{ALPL}_{i,m}^M =  \frac{1}{T-m-t_0+1}\sum_{t=t_0}^{T-m} \log p(y_{i,t+m}= y_{i,t+m}^{\text{o}}\gvn \by_{1:t}),
\]
where $p(y_{i,t+m}= y_{i,t+m}^{\text{o}}\gvn \by_{1:t})$ is the predictive likelihood. For this metric, a larger value indicates better forecast performance. 

To compare the forecast performance of model $M$ against the benchmark $B$, we follow \citet{CCM15} to report the percentage gains in terms of RMSFE, defined as
\[
	100 \times (1 - \text{RMSFE}_{i,m}^M/\text{RMSFE}_{i,m}^B),
\]
and the percentage gains in terms of ALPL:
\[
	100 \times (\text{ALPL}_{i,m}^M - \text{ALPL}_{i,m}^B).
\]

\begin{figure}[H]
    \centering
   \includegraphics[width=.85\textwidth]{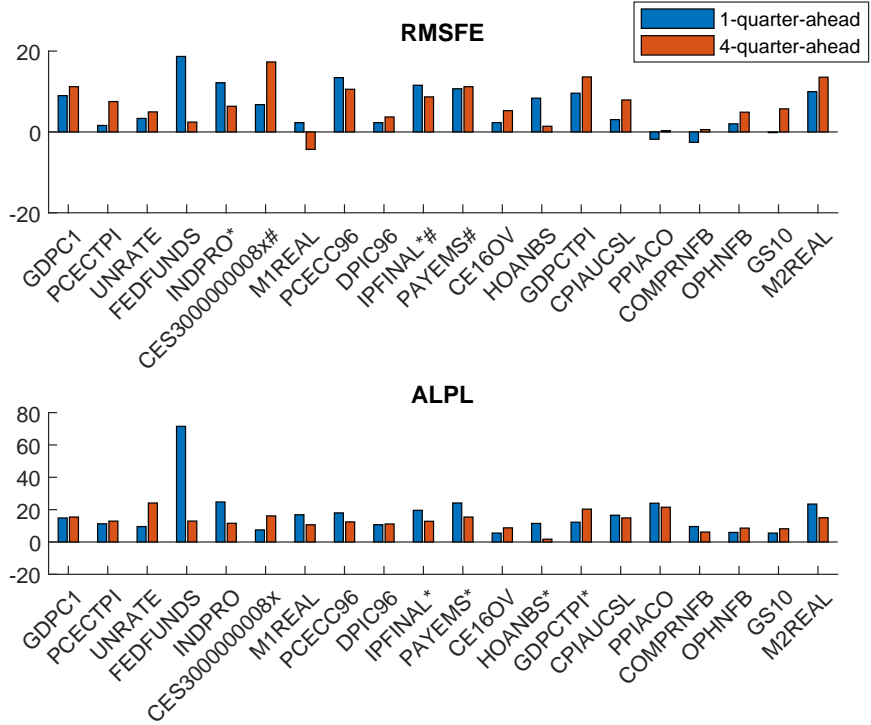}
   \caption{Forecasting results of the hybrid TVP-VAR compared to the benchmark: a standard homoscedastic, constant-coefficient VAR. The top and bottom panels show, respectively, the percentage gains in the root mean squared forecast error	and the average of log predictive likelihoods of the proposed hybrid TVP-VAR. The symbols * and \# after the mnemonic indicate rejection of equal forecast accuracy at significance level 0.05 using the test in \citet{DM95} for 1- and 4-quarter-ahead point forecasts, respectively.}
   \label{fig:forecasts_HYBvsBVAR}	
\end{figure}

Figure~\ref{fig:forecasts_HYBvsBVAR}	reports the forecasting results of the hybrid TVP-VAR, where we use a conventional homoscedastic, constant-coefficient VAR as the benchmark. The top panel shows the percentage gains in RMSFE for all 20 variables, and the bottom panel presents the corresponding results in ALPL. 

For both 1- and 4-quarter-ahead point forecasts, the hybrid TVP-VAR outperforms the benchmark for almost all variables (all but two for 1-quarter-ahead and one for 4-quarter ahead). For a few variables, such as the federal funds rate, real personal consumption expenditure and industrial production, the hybrid TVP-VAR outperforms the benchmark by more than 10\% for 1-quarter-ahead forecasts (the differences in forecast performance are also statistically significant at 0.05 level according to the test of \citet{DM95}). Overall, the median percentage gains in RMSFE for 1- and 4-quarter-ahead forecasts are, respectively, 5.1\% and 6.0\%.

For density forecasts, the hybrid TVP-VAR performs even better relative to the benchmark---it outperforms the benchmark for all variables in both forecast horizons. The median percentage gains in ALPL for 1- and 4-quarter-ahead forecasts are 14\% and 13\%, respectively. Moreover, for many variables the percentage gains are more than 20\%. These results are consistent with numerous studies in the small VAR literature, such as \citet{clark11}, \citet*{DGG13} and \citet{CR15}, that show allowing for time-varying structures substantially improves forecast performance compared to VARs with constant parameters, especially for density forecasts.

\begin{figure}[H]
    \centering
   \includegraphics[width=.85\textwidth]{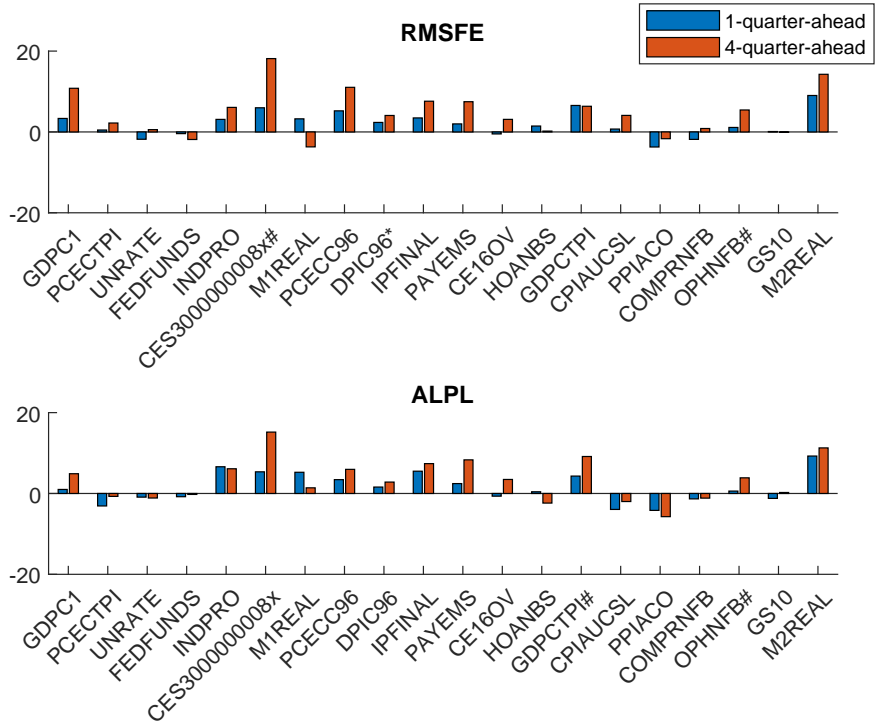}
   \caption{Forecasting results of the hybrid TVP-VAR compared to the benchmark HYB-$(0,0)$. The top and bottom panels show, respectively, the percentage gains in the root mean squared forecast error	and the average of log predictive likelihoods of the proposed hybrid TVP-VAR. The symbols * and \# after the mnemonic indicate rejection of equal forecast accuracy at significance level 0.05 using the test in \citet{DM95} for 1- and 4-quarter-ahead point forecasts, respectively.}
   \label{fig:forecasts_HYBvsHYB00}	
\end{figure}

Next, we compare the forecast performance of the hybrid TVP-VAR with that of HYB-$(0,0)$, the constant-coefficient VAR with stochastic volatility. The results are reported in Figure~\ref{fig:forecasts_HYBvsHYB00}. For both 1- and 4-quarter-ahead point forecasts, the hybrid TVP-VAR outperforms the HYB-$(0,0)$ for most variables. The median percentage gains in RMSFE are 1.7\% and 4.1\%, respectively. For density forecasts, the results are similar: the median percentage gains in ALPL for 1- and 4-quarter-ahead forecasts are, respectively, 0.8\% and 3.1\%. Overall, these results suggest that allowing for time variation in VAR coefficients---with appropriate shrinkage and sparsification---can further enhance the forecast performance of a VAR with stochastic volatility.

\begin{figure}[H]
    \centering
   \includegraphics[width=.85\textwidth]{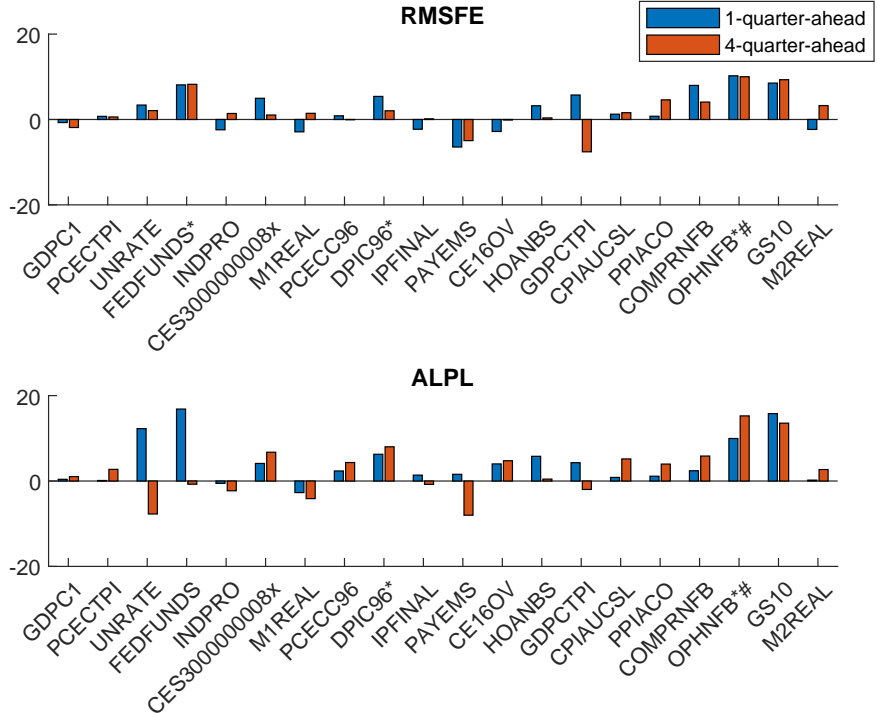}
   \caption{Forecasting results of the hybrid TVP-VAR compared to the benchmark HYB-$(1,1)$. The top and bottom panels show, respectively, the percentage gains in the root mean squared forecast error	and the average of log predictive likelihoods of the proposed hybrid TVP-VAR. The symbols * and \# after the mnemonic indicate rejection of equal forecast accuracy at significance level 0.05 using the test in \citet{DM95} for 1- and 4-quarter-ahead point forecasts, respectively.}
   \label{fig:forecasts_HYBvsHYB11}	
\end{figure}

Finally, Figure~\ref{fig:forecasts_HYBvsHYB11} compares the forecast performance of the 
the hybrid TVP-VAR with that of HYB-$(1,1)$, the full-fledged TVP-VAR where all the VAR coefficients and error variances are time varying. Again, for both point and density forecasts, the hybrid TVP-VAR performs better than the benchmark for most variables. In particular, the median percentage gains in RMSFE for 1- and 4-quarter-ahead forecasts are 1.0\% and 1.4\%, respectively; the median percentage gains in ALPL are 2.4\% and 2.8\%, respectively. These results suggest that imposing time variation in all equations is not necessary and would adversely impact the forecast performance. 

Overall, these forecasting results show that the proposed hybrid TVP-VAR forecasts better than many state-of-the-art time-varying models. These forecasting results highlight the advantages of using a data-driven approach to discover the time-varying structures---rather than imposing either constant coefficients or time variation in parameters.

\begin{figure}[H]
    \centering
   \includegraphics[width=.85\textwidth]{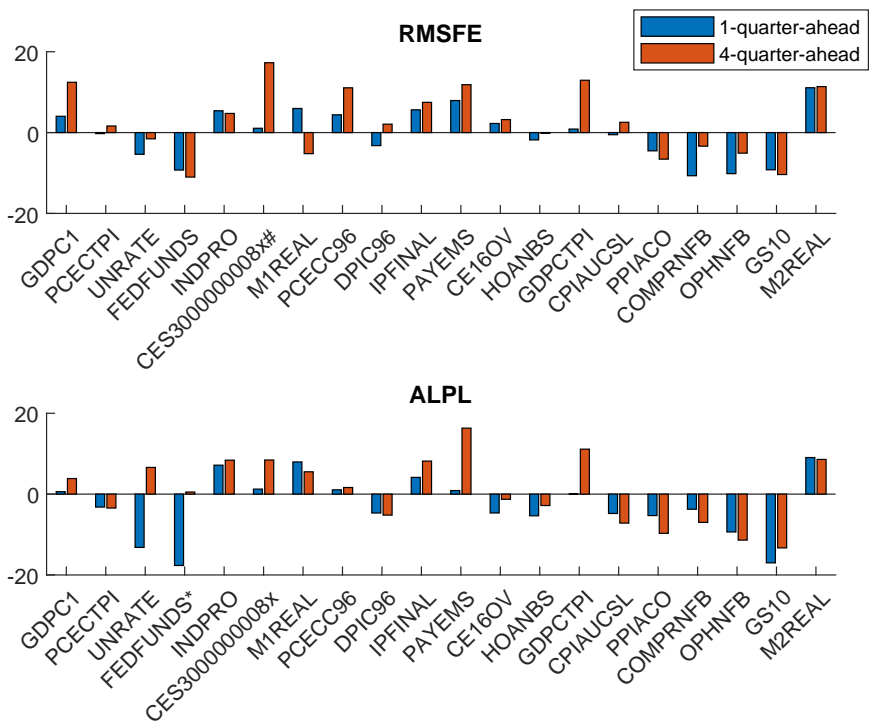}
   \caption{Forecasting results of HYB-$(1,1)$ compared to the benchmark HYB-$(0,0)$. The top and bottom panels show, respectively, the percentage gains in the root mean squared forecast error	and the average of log predictive likelihoods of the proposed hybrid TVP-VAR. The symbols * and \# after the mnemonic indicate rejection of equal forecast accuracy at significance level 0.05 using the test in \citet{DM95} for 1- and 4-quarter-ahead point forecasts, respectively.}
   \label{fig:forecasts_HYB11vsHYB00}	
\end{figure}

To better understand the sources of forecast gains, Figure~\ref{fig:forecasts_HYB11vsHYB00} compares the forecast performance of HYB-$(1,1)$, the full-fledged TVP-VAR, to HYB-$(0,0)$, the constant-coefficient VAR with stochastic volatility. The results are mixed: while HYB-$(1,1)$ does slightly better in terms of point forecasts for the majority of the variables, it performs worse in terms of density forecasts for many variables. These results suggest that allowing for time-varying VAR coefficients in all equations does not necessarily improve forecast performance. (The forecast performance of HYB-$(1,1)$ and HYB-$(1,0)$ are very similar, as reported in Appendix D). This finding is also consistent with the full-sample estimation results presented in Table~\ref{tab:gam_n20}: while the data clearly favors time-varying VAR coefficients in a few equations, for the majority of the equations time variation is not needed.

Finally, one can also interpret the superior forecast performance of the hybrid TVP-VAR through the lens of Bayesian forecast combinations \citep{MZ93, AK08}. As discussed in Section~\ref{ss:BMA}, the proposed hybrid TVP-VAR can be viewed as a Bayesian model average of a wide variety TVP-VARs with different forms of time variation, where each component model is characterized by the vector of indicators $\vgamma$. Then, the forecasts from the hybrid TVP-VAR can be interpreted as a forecast combination weighted by the posterior model probabilities $p(\vgamma\gvn \by)$. Consistent with the large literature on forecast combinations, here we find that the forecast combination of the hybrid TVP-VAR performs better than many individual component models, including HYB-$(1,1)$ and HYB-$(0,0)$.

\section{Concluding Remarks and Future Research} \label{s:conclusions}

This paper has developed what we call hybrid TVP-VARs, i.e., VARs with time-varying parameters in some equations but not in others. Using US data, we found evidence that while VAR coefficients and error covariances in some equations are time varying, the data prefers constant coefficients in others. In a forecasting exercise that involves 20 macroeconomic and financial variables, we demonstrated the superior forecast performance of the proposed hybrid TVP-VARs compared to standard benchmarks. 

In future work, it would be interesting to use hybrid TVP-VARs for structural analysis. Since they are formulated in the recursive structural-form, structural analysis using large VARs identified by recursive zero restrictions, such as the application in \citet{ER17}, can directly use the proposed models. 

In addition, developing an order-invariant version of these hybrid TVP-VARs would be an interesting and important extension. This would involve changing two components. First, one requires the use of the reduced-form VAR representation, but model indicators $\vgamma$ can be introduced similarly. An equation-by-equation estimation procedure can be developed along the lines in \citet{CCCM21}, though computation would be more intensive. Second, one would need to replace the Cholesky stochastic volatility model with an order-invariant model suitable for large VARs, such as \citet{CCM16} or \citet{CKY21}. The trade-off between the two stochastic volatility models is between computational speed and model flexibility: the former can be estimated quickly but the latter is more flexible. Finding a good modeling approach among all these choices---with an eye on the additional computational costs---would be an interesting research direction. 

\newpage

\section*{Appendix A: Estimation Details}

In this appendix we provide estimation details of the hybrid TVP-VAR given in \eqref{eq:yi}-\eqref{eq:hi}.
In particular, we describe the details of Step 2 - Step 7 of the posterior sampler outlined in Section 3.2 of the main text.

\underline{\textbf{Step 2}}. To sample $\bh_{i}$, the log-volatility vector of the $i$-th equation, 
$i=1,\ldots, n,$ we can use the auxiliary mixture sampler of \citet{KSC98}. More specifically, we first compute the residuals $\epsilon_{i,t}^y, t=1,\ldots, T,$ using \eqref{eq:yi}. Then, we transform these residuals as $\by_i^* = (\log (\epsilon_{i,1}^y)^2,\ldots, \log (\epsilon_{i,T}^y)^2)'$. Finally, we implement the auxiliary mixture sampler in conjunction with the precision sampler of \citet{CJ09} to sample $\bh_{i}$ using $\by_i^*$ as data.

\underline{\textbf{Step 3}}. The parameters $\vSigma_{\theta_i}^{\frac{1}{2}} =\text{diag}(\vSigma_{\beta_i}^{\frac{1}{2}}, \vSigma_{\alpha_i}^{\frac{1}{2}})$ and $\vtheta_{i,0}$ can be sampled easily as their joint distribution is Gaussian. To see that, let $\vmu_i^\theta = (\vtheta_{i,0}',\sigma_{\beta_i,1},\ldots,\sigma_{\beta_i,k_{\beta}},\sigma_{\alpha_i,1},\ldots,\sigma_{\alpha_i,k_{\alpha_i}})'$ and define $\bw_{i,t}^{\theta} = (\bx_{i,t}, \gamma_i^\beta\tilde{\bx}_{t}\odot \tilde{\vbeta}_{i,t}, \gamma_i^\alpha\tilde{\bw}_{i,t}\odot \tilde{\valpha}_{i,t})$, where $\odot$ denotes the component-wise product. Then, we can rewrite \eqref{eq:yi} as a linear regression:
\[
	y_{i,t} = \bw_{i,t}^{\theta}  \vmu_i^\theta + \epsilon_{i,t}^y.
\]
Since both $\vSigma_{\theta_i}^{\frac{1}{2}}$ and $\vtheta_{i,0}$ have Gaussian priors, the implied prior on $\vmu_i^\theta$ is also Gaussian: $\vmu_i^\theta\sim\distn{N}(\mathbf{0},\bV_{\vmu_i^\theta})$, where $\bV_{\vmu_i^\theta} = \text{diag}(\bV_{\vtheta_{i,0}},S_{\theta_i,1},\ldots, S_{\theta_i,k_{\theta_i}})$. Define $\bW_i^{\theta}$ by stacking $\bw_{i,t}^\theta$ over 
$t=1,\ldots, T$. It follows that the full conditional distribution of $\vmu_i^\theta$ is given by
\[
	(\vmu_i^\theta \gvn  \by_i, \tilde{\vtheta}_i, \bh_i, \vgamma_i) \sim \distn{N}(\hat{\vmu}_i^\theta, \bK_{\vmu_i^\theta}^{-1}),
\]
where $\bK_{\vmu_i^\theta} = \bV_{\vmu_i^\theta}^{-1} + (\bW_i^{\theta})'\vOmega_{\bh_i}^{-1}\bW_i^{\theta} $ and 
$\hat{\vmu}_i^\theta = \bK_{\vmu_i^\theta}^{-1}(\bW_i^{\theta})'\vOmega_{\bh_i}^{-1}\by_i $ with $\vOmega_{\bh_i}=\text{diag}(\e^{h_{i,1}},\ldots,\e^{h_{i,T}})$.

\underline{\textbf{Steps 4-5}}. The full conditional distributions of $\sigma_{h,i}^2$ and $h_{i,0}$ are standard and 
they can be sampled easily. In particular, their full conditional distributions are
\begin{align*} 
	(\sigma_{h,i}^2 \gvn \bh_i, h_{i,0}) &\sim \distn{IG}\left(\nu_{h,i}+\frac{T}{2}, 
	S_{h,i} + \frac{1}{2}\sum_{t=1}^T(h_{i,t}-h_{i,t-1})^2\right), \\
	(h_{i,0} \gvn \bh_i, \sigma_{h,i}^2 ) & \sim \distn{N}(\hat{h}_{i,0}, K_{h_{i,0}}^{-1}),
\end{align*}
where $ K_{h_{i,0}} = 1/V_{h_{i,0}} + 1/\sigma_{h,i}^2 $ and $ \hat{h}_{i,0} = K_{h_{i,0}}^{-1}(a_{h_{i,0}}/V_{h_{i,0}} + h_{i,1}/\sigma_{h,i}^2)$.

\underline{\textbf{Step 6}}. Next, given the independent beta priors on $p^{\beta}_i$ and $p^{\alpha}_i, i=1,\ldots, n$  their full conditional posterior distributions are also beta distributions. In fact, we have:
\begin{align*}
	(p^{\beta}_i \gvn \gamma^{\beta}_i) \sim\distn{B}(a_{p^\beta} + \gamma^{\beta}_i,
	b_{p^\beta} + 1 - \gamma^{\beta}_i), \\
	(p^{\alpha}_i\gvn \gamma^{\alpha}_i) \sim\distn{B}(a_{p^{\alpha}} + \gamma^{\alpha}_i, 
	b_{p^{\alpha}} + 1-\gamma^{\alpha}_i).
\end{align*}

\underline{\textbf{Step 7}}. To implement Step 7, we follow the sampling approach in \citet{chan21}. First note that $\kappa_1$ and $\kappa_2$ only appear in their priors $\kappa_j\sim \distn{G}(c_{1,j},c_{2,j}), j=1,2$, and in the prior covariance matrices $\bV_{\vtheta_{i,0}}, i=1,\ldots, n$.  Letting $\theta_{ij,0}$ denote the $j$-th element of $\vtheta_{i,0}$, we define the index set $S_{\kappa_1}$ to be the collection of indexes $(i,j)$ such that $\theta_{ij,0}$ is a coefficient associated with an own lag. That is, $S_{\kappa_1} = \{(i,j): \theta_{ij,0} \text{ is a coefficient associated with an own lag}\}$. Similarly, define $S_{\kappa_2}$ as the set that collects all the indexes $(i,j)$ such that $\theta_{ij,0}$ is a coefficient associated with a lag of other variables. It is easy to check that the numbers of elements in $S_{\kappa_1}$  and $S_{\kappa_2}$ are respectively $np$ and $(n-1)np$. Furthermore, for $(i,j)\in S_{\kappa_1}\cup S_{\kappa_2}$, let
\[
	C_{ij} = \left\{
	\begin{array}{ll}
			\frac{1}{l^2}, & \text{for the coefficient on the $l$-th lag of variable } i,\\
			\frac{s_i^2}{l^2 s_j^2}, & \text{for the coefficient on the $l$-th lag of variable } j, j\neq i. \\			
	\end{array} \right.
\]
Then, we have
\begin{align*}
	p(\kappa_1 \gvn \vtheta_{0})	& \propto \prod_{(i,j)\in S_{\kappa_1}} \kappa_1^{-\frac{1}{2}} \e^{-\frac{1}{2\kappa_{1}C_{ij}}
	\theta_{ij,0}^2}\times \kappa_1^{c_{1,1}-1}\e^{-\kappa_1 c_{2,1}} \\
	 & = \kappa_1^{c_{1,1}-\frac{np}{2}-1} \e^{-\frac{1}{2} \left(2c_{2,1}\kappa_1 + \kappa_{1}^{-1}\sum_{(i,j)\in S_{\kappa_1}}\frac{\theta_{ij,0}^2}{C_{ij}}\right)},
\end{align*}
which is the kernel of the generalized inverse Gaussian distribution:
\[
	(\kappa_1 \gvn \vtheta_{0}) \sim \distn{GIG}\left(c_{1,1}-\frac{np}{2}, 2c_{2,1}, \sum_{(i,j)\in S_{\kappa_1}}\frac{\theta_{ij,0}^2}{C_{ij}}\right).
\]
Similarly, $\kappa_2$ also has a generalized inverse Gaussian distribution:
\[
	(\kappa_2 \gvn \vtheta_{0}) \sim \distn{GIG}\left(c_{1,2}-\frac{(n-1)np}{2}, 2c_{2,2},\sum_{(i,j)\in S_{\kappa_2}} \frac{\theta_{ij,0}^2}{C_{ij}}\right).
\]

\newpage

\section*{Appendix B: Technical Details on Model Comparison}

This appendix outlines the technical details on comparing hybrid TVP-VARs using the Savage-Dickey density ratio. First, suppose we want to compare the proposed hybrid TVP-VAR with $\vgamma$ unrestricted against a TVP-VAR characterized by $\vgamma = \bc\in\{0,1\}^{2n}$. Then, the Bayes factor in favor of the unrestricted model can be expressed as the Savage-Dickey density ratio
\[
	\text{BF}_{\text{u},\bc} = \frac{p(\vgamma=\bc)}{p(\vgamma=\bc\gvn \by)},
\]
provided that the priors under the restricted and unrestricted models satisfy a compatibility condition \citep[see, e.g.,][for details]{VW95}. The priors described in Section~\ref{ss:priors} of the main text assume that the restricted (i.e., $\vgamma$) and unrestricted parameters are independent \textit{a priori}, which implies the compatibility condition.

Next, we outline how one can evaluate the marginal prior and posterior densities at $\vgamma =\bc$. First, to evaluate $p(\vgamma=\bc)$, recall that under our priors, $\gamma^\beta_i$ and $\gamma^\alpha_i$ follow independent Bernoulli distributions with success probabilities $p^{\beta}_i$ and $p^{\alpha}_i$, respectively, for $i=1,\ldots, n$. These success probabilities, in turn, are assumed to have beta distributions: $p^{\beta}_i \sim\distn{B}(a_{p^\beta},b_{p^\beta})$ and $p^{\alpha}_i\sim\distn{B}(a_{p^{\alpha}},b_{p^{\alpha}})$. Hence, the marginal prior of $\gamma^\beta_i$ (unconditional on $p^{\beta}_i$) can be computed as
\begin{align*}
	p(\gamma_i^\beta) & = \int_{0}^1 (p^{\beta}_i)^{\gamma_i^\beta}(1-p^{\beta}_i)^{1-\gamma_i^\beta}
	\times \frac{\Gamma(a_{p^\beta}+b_{p^\beta})}{\Gamma(a_{p^\beta})\Gamma(b_{p^\beta})} 
	(p^{\beta}_i)^{a_{p^\beta}-1}(1-p^{\beta}_i)^{a_{p^\beta}-1}\di p^{\beta}_i \\
	& = \frac{\Gamma(a_{p^\beta}+b_{p^\beta})\Gamma(\gamma_i^{\beta}+a_{p^\beta})\Gamma(1-\gamma_i^{\beta}+b_{p^\beta})}
	{\Gamma(a_{p^\beta})\Gamma(b_{p^\beta})\Gamma(a_{p^\beta}+b_{p^\beta}+1)}, 
\end{align*}
where $\Gamma(\cdot)$ is the gamma function. A similar expression can be derived for $p(\gamma_i^\alpha)$. It follows that the marginal prior for $\vgamma$ has the following analytical expression
\begin{align*}
	p(\vgamma) & = \prod_{i=1}^n p(\gamma_i^\beta)p(\gamma_i^\alpha) \\
	& = \prod_{i=1}^n \frac{\Gamma(a_{p^\beta}+b_{p^\beta})\Gamma(\gamma_i^{\beta}+a_{p^\beta})\Gamma(1-\gamma_i^{\beta}+b_{p^\beta})} {\Gamma(a_{p^\beta})\Gamma(b_{p^\beta})\Gamma(a_{p^\beta}+b_{p^\beta}+1)}
	\frac{\Gamma(a_{p^\alpha}+b_{p^\alpha})\Gamma(\gamma_i^{\alpha}+a_{p^\alpha})\Gamma(1-\gamma_i^{\alpha}+b_{p^\alpha})} {\Gamma(a_{p^\alpha})\Gamma(b_{p^\alpha})\Gamma(a_{p^\alpha}+b_{p^\alpha}+1)},
\end{align*}
which can be easily evaluated at any point $\bc$. 

Next, we can evaluate $p(\vgamma=\bc\gvn \by)$ with $\bc = (c_{1,1},c_{1,2},\ldots, c_{n,1}, c_{n,2})'$ using the Monte Carlo average:
\[
	\frac{1}{R}\sum_{r=1}^R \prod_{i=1}^n \Pm\left(\vgamma_i = (c_{i,1},c_{i,2}) \gvn  \by_i, \bh_i^{(r)}, \vSigma_{\theta_i}^{(r)}, \vtheta_{i,0}^{(r)}\right),	
\]
where $\bh_i^{(r)}, \vSigma_{\theta_i}^{(r)}, \vtheta_{i,0}^{(r)}, i=1,\ldots, n, r=1,\ldots, R$ are posterior draws from the unrestricted model. Analytical expressions of the above posterior probabilities are given in Section~\ref{ss:estimation} of the main text. For numerical stability, both the prior and posterior densities are computed in log scale. 

Thus, we have shown how one can compare the proposed hybrid TVP-VAR with $\vgamma$ unrestricted against any TVP-VAR
characterized by $\vgamma = \bc\in\{0,1\}^{2n}$. To compare two TVP-VARs characterized by $\vgamma = \bc_1$ and $\vgamma = \bc_2$, note that one can express the Bayes factor in favor of $\vgamma = \bc_1$ as
\[
	\text{BF}_{\bc_1,\bc_2} = \frac{\text{BF}_{\text{u},\bc_2}}{\text{BF}_{\text{u},\bc_1}} = 
	\frac{p(\vgamma=\bc_1\gvn \by)}{p(\vgamma=\bc_2\gvn \by)}.
\]
This expression can be evaluated using posterior draws from the unrestricted model as before.

\newpage

\section*{Appendix C: Data}

The dataset covers 20 quarterly variables sourced from the FRED-QD database at the Federal Reserve Bank of St. Louis \citep{MN21}. The sample period is from 1959Q1 to 2018Q4.
Table~\ref{tab:var} lists all the variables and describes how they are transformed. For example, $\Delta \log $ is used to denote 
the first difference in the logs, i.e., $\Delta \log x = \log x_t - \log x_{t-1}$.

\begin{table}[H]
\centering
\caption{Description of variables used in empirical application.} \label{tab:var}
\resizebox{\textwidth}{!}{
\begin{tabular}{lll}
\hline\hline
Variable & Mnemonic & Transformation \\ \hline
Real Gross Domestic Product 	&	GDPC1	&	400$\Delta \log$	\\
\rowcolor{lightgray}
Personal Consumption Expenditures: Chain-type & & \\
\rowcolor{lightgray}
Price index	&	PCECTPI	&	400$\Delta \log$	\\
Civilian Unemployment Rate	&	UNRATE	&	no transformation	\\
\rowcolor{lightgray}
Effective Federal Funds Rate	&	FEDFUNDS	&	no transformation	\\
Industrial Production Index	&	INDPRO	&	400$\Delta \log$	\\
\rowcolor{lightgray}
Real Average Hourly Earnings of Production and & & \\
\rowcolor{lightgray}
Nonsupervisory  Employees: Manufacturing	&	CES3000000008x	&	400$\Delta \log$	\\
Real M1 Money Stock	&	M1REAL	&	400$\Delta \log$	\\
\rowcolor{lightgray}
Real Personal Consumption Expenditures	&	PCECC96	&	400$\Delta \log$	\\
Real Disposable Personal Income	&	DPIC96	&	400$\Delta \log$	\\
\rowcolor{lightgray}
Industrial Production: Final Products	&	IPFINAL	&	400$\Delta \log$	\\
All Employees: Total nonfarm	&	PAYEMS	&	400$\Delta \log$	\\
\rowcolor{lightgray}
Civilian Employment	&	CE16OV	&	400$\Delta \log$	\\
Nonfarm Business Section: Hours of All Persons	&	HOANBS	&	400$\Delta \log$	\\
\rowcolor{lightgray}
Gross Domestic Product: Chain-type Price index	&	GDPCTPI	&	400$\Delta \log$	\\
Consumer Price Index for All Urban Consumers: All Items	&	CPIAUCSL	&	400$\Delta \log$	\\
\rowcolor{lightgray}
Producer Price Index for All commodities	&	PPIACO	&	400$\Delta \log$	\\
Nonfarm Business Sector: Real Compensation Per Hour	&	COMPRNFB	&	400$\Delta \log$	\\
\rowcolor{lightgray}
Nonfarm Business Section: Real Output Per Hour of & & \\
\rowcolor{lightgray}
All Persons	&	OPHNFB	&	400$\Delta \log$	\\
10-Year Treasury Constant Maturity Rate	&	GS10	&	no transformation	\\
\rowcolor{lightgray}
Real M2 Money Stock	&	M2REAL	&	400$\Delta \log$	\\
\hline\hline
\end{tabular}
}
\end{table}

\newpage

\section*{Appendix D: Additional Results}

In this appendix we provide additional simulation and empirical results. First, we investigate the effect of the prior on $\gamma_i^{\beta}$ and $\gamma_i^{\alpha}$ by replacing the beta prior $\distn{B}(0.5,0.5)$ by the uniform prior $\distn{U}(0,1)$ in the Monte Carlo experiments. Table~\ref{tab:gam_2} reports the frequencies of the posterior modes of $\gamma^\beta_i$ and $\gamma^\alpha_i$ being one in 300 datasets. All in all, the simulation results are similar to the baseline case, suggesting that the posterior estimates are not sensitive to the prior on $\gamma_i^{\beta}$ and $\gamma_i^{\alpha}$.

\begin{table}[H]
\caption{Frequencies (\%) of the posterior modes of $\gamma^\beta_i$ and $\gamma^\alpha_i$ being one in 300 datasets with $T=400$. The two sets of priors for $\gamma_i^{\beta}$	and $\gamma_i^{\alpha}$ are 
$\distn{B}(0.5,0.5)$ and $\distn{U}(0,1)$ priors.}
\label{tab:gam_2}
\centering
\begin{tabular}{ccccccc}\hline\hline
Equation	&	True $\gamma_i^{\beta}$	&	True $\gamma_i^{\alpha}$	&	\multicolumn{2}{c}{$\distn{B}(0.5,0.5)$ prior}		&	\multicolumn{2}{c}{$\distn{U}(0,1)$ prior}			\\
	&		&		&	$\gamma_i^{\beta}$	&	$\gamma_i^{\alpha}$ 	&	$\gamma_i^{\beta}$	&	$\gamma_i^{\alpha}$ 	\\ \hline
1	&	0	&	0	&	0.06	&	--	&	0.06	&	--	\\
\rowcolor{lightgray}
2	&	0	&	1	&	0.04	&	0.88	&	0.03	&	0.87	\\
3	&	1	&	0	&	0.98	&	0.25	&	0.99	&	0.24	\\
\rowcolor{lightgray}
4	&	1	&	1	&	0.98	&	0.64	&	0.98	&	0.62	\\
5	&	0	&	0	&	0.02	&	0.02	&	0.00	&	0.02	\\
\rowcolor{lightgray}
6	&	0	&	1	&	0.03	&	0.96	&	0.02	&	0.96	\\
7	&	1	&	0	&	0.97	&	0.13	&	0.98	&	0.11	\\
\rowcolor{lightgray}
8	&	1	&	1	&	0.95	&	0.80	&	0.95	&	0.81	\\
9	&	0	&	0	&	0.03	&	0.00	&	0.01	&	0.00	\\
\rowcolor{lightgray}
10	&	0	&	1	&	0.04	&	0.94	&	0.02	&	0.96	\\
11	&	1	&	0	&	0.94	&	0.11	&	0.97	&	0.10	\\
\rowcolor{lightgray}
12	&	1	&	1	&	0.93	&	0.88	&	0.93	&	0.90	\\ \hline \hline
\end{tabular}
\end{table}

\begin{figure}[H]
    \centering
   \includegraphics[width=.65\textwidth]{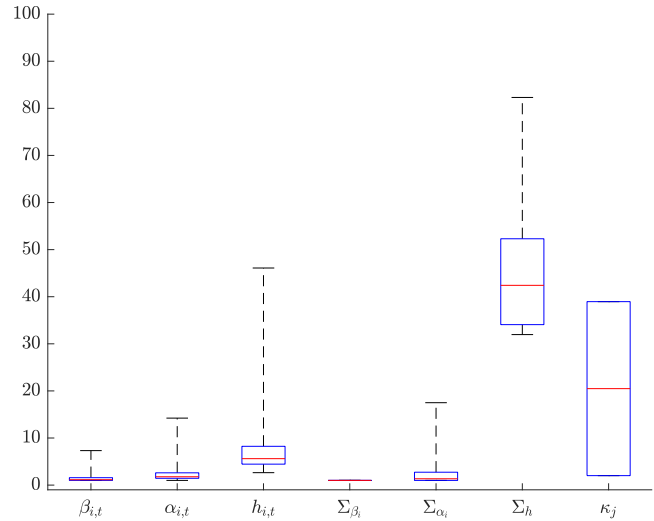}
   \caption{Boxplots of inefficiency factors of posterior draws from the 20-variable hybrid TVP-VAR.}
   \label{fig:IF}	
\end{figure}

Next, we compute the inefficiency factors of the posterior draws from the 20-variable hybrid TVP-VAR in Section~\ref{s:application} of the main text, defined as
\[
	1 + 2\sum_{l=1}^L \rho_l, 
\]
where $\rho_l$ is the sample autocorrelation at lag length $l$ and $L$ is chosen to be large enough so that the autocorrelation tapers off. In the ideal case where the posterior draws are independent, the corresponding inefficiency factor is 1. Figure~\ref{fig:IF} reports the inefficiency factors, obtained using 10,000 posterior draws after a burn-in period of 1,000. The results show that the proposed posterior sampler is efficient in terms of producing posterior draws that are not highly autocorrelated.

\begin{figure}[H]
    \centering
   \includegraphics[width=.85\textwidth]{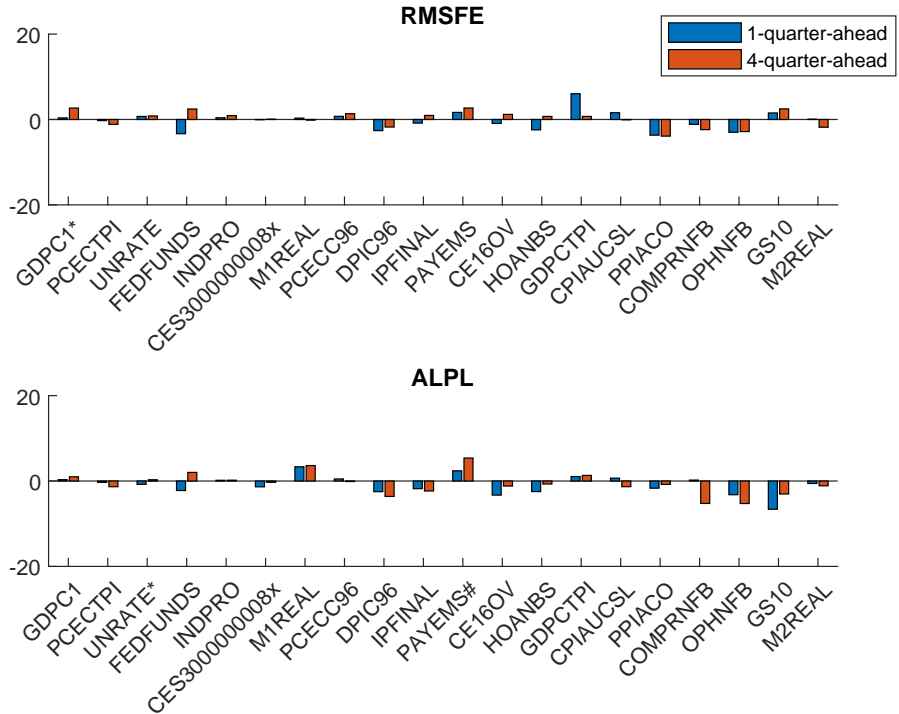}
   \caption{Forecasting results of HYB-$(1,1)$ compared to the benchmark HYB-$(1,0)$. The top and bottom panels show, respectively, the percentage gains in the root mean squared forecast error	and the average of log predictive likelihoods of the proposed hybrid TVP-VAR. The symbols * and \# after the mnemonic indicate rejection of equal forecast accuracy at significance level 0.05 using the test in \citet{DM95} for 1- and 4-quarter-ahead point forecasts, respectively.}
   \label{fig:forecasts_HYB11vsHYB10}	
\end{figure}

Finally, we report additional forecasting results. First, Figure~\ref{fig:forecasts_HYB11vsHYB10} reports forecasting results of the full-fledged TVP-VAR  HYB-$(1,1)$ relative to the benchmark HYB-$(1,0)$, a TVP-VAR with stochastic volatility and a constant impact matrix (by setting all $\gamma_i^{\beta}$ to 1 and all $\gamma_i^{\alpha}$ to 0). The results show that the forecast performance of the two models are mostly similar.

\begin{figure}[H]
    \centering
   \includegraphics[width=.85\textwidth]{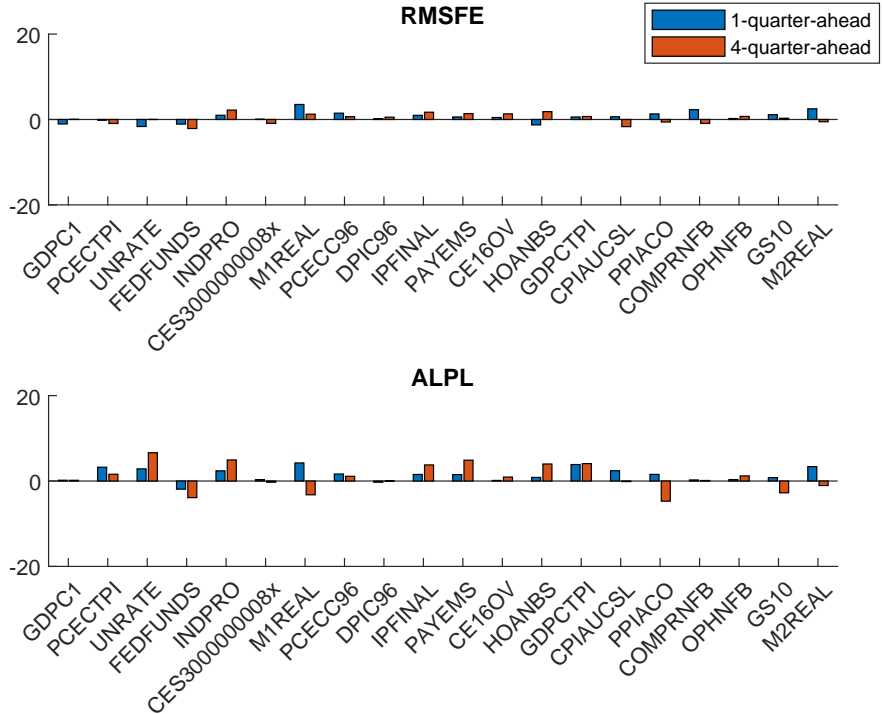}
   \caption{Forecasting results of an extension of HYB where the innovations to $h_{i,t}$ are correlated across equations compared to the benchmark HYB. The top and bottom panels show, respectively, the percentage gains in the root mean squared forecast error	and the average of log predictive likelihoods of the proposed hybrid TVP-VAR. The symbols * and \# after the mnemonic rejection of equal forecast accuracy at significance level 0.05 using the test in \citet{DM95} for 1- and 4-quarter-ahead point forecasts, respectively.}
   \label{fig:forecasts_HYBcorrhvsHYB}	
\end{figure}

Next, Figure~\ref{fig:forecasts_HYBcorrhvsHYB} reports the forecasting results of an extension of the hybrid TVP-VAR where the innovations to $h_{i,t}$ are correlated across equations relative to the benchmark HYB where the innovations are independent. The results show that the forecast performance of the extension is on average better in both point and density forecasts, although the forecast gains are modest. For example, the median percentage gains in RMSFE for 1- and 4-quarter-ahead forecasts are 0.6\% and 0.4\%, respectively; the median percentage gains in ALPL are 1.5\% and 0.6\%, respectively.

\begin{figure}[H]
    \centering
   \includegraphics[width=.8\textwidth]{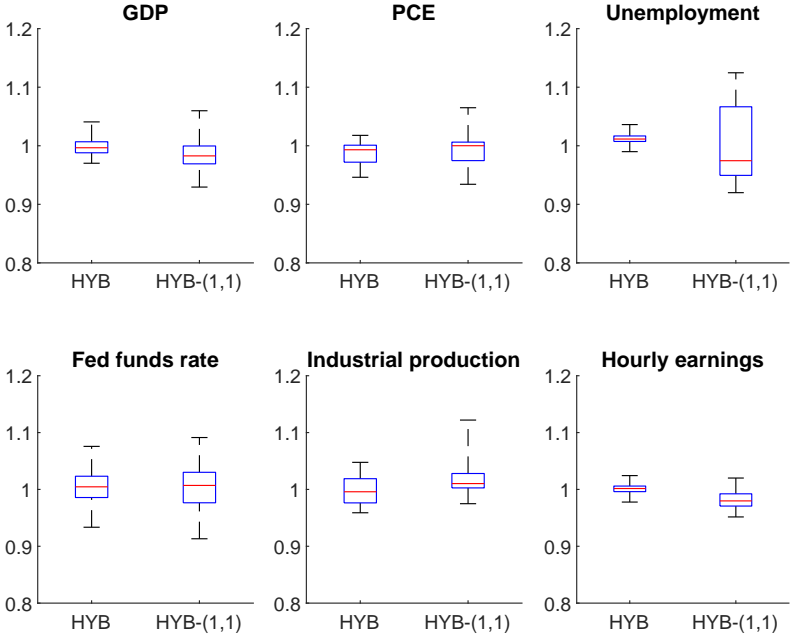}
   \caption{Boxplots of the relative mean squared forecast errors of 1-quarter-ahead point forecasts from the proposed hybrid TVP-VAR (HYB) and the full-fledge TVP-VAR (HYB-$(1,1)$).}
   \label{fig:ordering_forecasts_RMSFE1}	
\end{figure}

\begin{figure}[H]
    \centering
   \includegraphics[width=.8\textwidth]{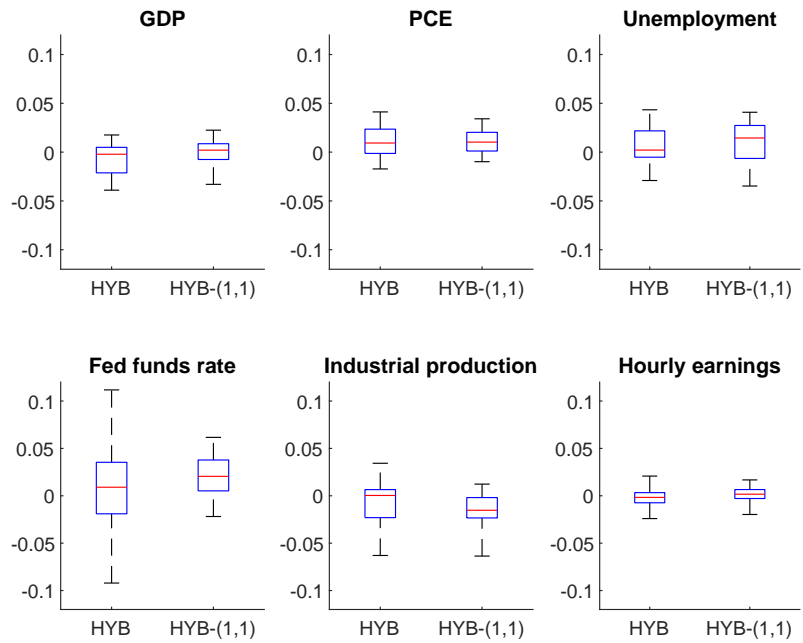}
   \caption{Boxplots of the relative average log predictive likelihoods of 1-quarter-ahead density forecasts from the proposed hybrid TVP-VAR (HYB) and the full-fledge TVP-VAR (HYB-$(1,1)$).}
   \label{fig:ordering_forecasts_ALPL1}	
\end{figure}

Next, we investigate the variability of point and density forecasts of 6 variables---real GDP, PCE inflation, unemployment, Fed funds rate and industrial production and real average hourly earnings in manufacturing---from the proposed model. We consider all 720 possible variable orderings and the setup of the forecasting exercise is described in Section 5.4 in the main text. For each variable ordering, we compute the root mean squared forecast error and the average of log predictive likelihoods from the proposed model as well as the HYB-$(1,1)$. We normalize the two metrics by those of the benchmark ordering, and the results are reported in Figures~\ref{fig:ordering_forecasts_RMSFE1}-\ref{fig:ordering_forecasts_ALPL1}.

Consistent with the results in \citet{ARRS21}, the variability of point forecasts is relatively small. Moreover, the variability of both point and density forecasts from the proposed model is similar to those in HYB-$(1,1)$, a structural-form parameterization of the model in \citet{Primiceri05}.

\newpage

\singlespace


\ifx\undefined\BySame
\newcommand{\BySame}{\leavevmode\rule[.5ex]{3em}{.5pt}\ }
\fi
\ifx\undefined\textsc
\newcommand{\textsc}[1]{{\sc #1}}
\newcommand{\emph}[1]{{\em #1\/}}
\let\tmpsmall\small
\renewcommand{\small}{\tmpsmall\sc}
\fi

\end{document}